\PassOptionsToPackage{pdfpagelabels=false}{hyperref} 
\documentclass[a4paper,fleqn,usenatbib,useAMS]{mnras}
\usepackage{xspace}

\usepackage{hyperref}

\usepackage{graphicx}
\usepackage{inputenc}
\usepackage[T1]{fontenc}
\usepackage{txfonts}
\usepackage{lmodern}
\usepackage{epsfig}
\usepackage{hyperref}
\usepackage{amsmath}

\usepackage{natbib}
\usepackage{relsize}
\usepackage{ae,aecompl}
\usepackage{graphicx}	% Including figure files
\usepackage{amssymb}	% Extra maths symbols
\usepackage{wasysym}
\usepackage{multicol}        % Multi-column entries in tables
\usepackage{bm}		% Bold maths symbols, including upright Greek
\usepackage{pdflscape}	% Landscape pages

\newcommand{\kms}{km~s$^{-1}$ }
\newcommand{\kmsp}{km~s$^{ -1}$}
\newcommand{\oi}{O~{\sc I}~}

\newcommand{\nv}{N~{\sc V}~}
\newcommand{\cii}{C~{\sc II}~}

\newcommand{\siii}{Si~{\sc II}~}
\newcommand{\siiii}{Si~{\sc III}~}
\newcommand{\sii}{S~{\sc II}~}
\newcommand{\siiifs}{Si~{\sc II$^\ast$}~}
\newcommand{\siiv}{Si~{\sc IV}~}

\newcommand{\oip}{O~{\sc I}}

\newcommand{\siiip}{Si~{\sc II}}
\newcommand{\siiifsp}{Si~{\sc II$^\ast$}}
\newcommand{\siiiip}{Si~{\sc III}}

\newcommand{\siivp}{Si~{\sc IV}}

\newcommand{\mout}{$\dot{M}_{\text{o}}$ }
\newcommand{\mstar}{$M_\ast$ }
\newcommand{\mstarp}{$M_\ast$}

\newcommand{\moutp}{$\dot{M}_{\text{o}}$}

\begin{document}

\title[The Mass Outflow Rate of NGC 6090]{A Robust Measurement of the Mass Outflow Rate of the Galactic Outflow from NGC 6090}
\author[Chisholm et al.]{John Chisholm$^{1}$\thanks{Contact email: chisholm@astro.wisc.edu}, Christy A. Tremonti$^{1}$, Claus Leitherer $^{2}$, Yanmei Chen$^{3}$\\
$^{1}$Astronomy Department, University of Wisconsin, Madison, 475
  N. Charter St., WI 53711, USA \\
$^{2}$Space Telescope Science Institute, 3700 San Martin Drive, Baltimore, MD 21218, USA \\
$^{3}$Department of Astronomy, Nanjing University, Nanjing 210093, China\\
}
\pubyear{2016}
\label{firstpage}
\pagerange{\pageref{firstpage}--\pageref{lastpage}}
\maketitle
\begin{abstract}
To evaluate the impact of stellar feedback, it is critical to estimate the mass outflow rates of galaxies.  Past estimates have been plagued by uncertain assumptions about the outflow geometry, metallicity, and ionization fraction.  Here we use Hubble Space Telescope ultraviolet spectroscopic observations of the nearby starburst NGC 6090 to demonstrate that many of these quantities can be constrained by the data. We use the Si~{\sc IV} absorption lines to calculate the scaling of velocity (v), covering fraction (C$_f$), and density with distance from the starburst (r), assuming the Sobolev optical depth and a velocity law of the form:  v$~\propto(1 -\mathrm{R}_\mathrm{i}/\mathrm{r} )^\beta$ (where R$_\mathrm{i}$ is the inner outflow radius).  We find that the velocity ($\beta$=0.43) is consistent with an outflow driven by an r$^{-2}$ force with the outflow radially accelerated, while the scaling of the covering fraction ($C_f \propto \mathrm{r}^{-0.82}$) suggests that cool clouds in the outflow are in pressure equilibrium with an adiabatically expanding medium. We use the column densities of four weak metal lines and CLOUDY photoionization models to determine the outflow metallicity, the ionization correction, and the initial density of the outflow. Combining these values with the profile fitting, we find R$_\mathrm{i}$ = 63 pc, with most of the mass within 300~pc of the starburst. Finally, we find that the maximum mass outflow rate is 2.3~M$_\odot$ yr$^{-1}$ and the mass loading factor (outflow divided by the star formation rate) is 0.09, a factor of 10 lower than the value calculated using common assumptions for the geometry, metallicity and ionization structure of the outflow.
\end{abstract}

\section{INTRODUCTION}

Strangely, most of the gas within a galaxy is not near stars \citep{songaila, adelberger, tumlinson, werk, peeples14, wakker2015}. The circum-galactic medium extends to radii greater than 150~kpc, is metal rich, and spans a range of temperatures \citep{songaila, werk, peeples14}. Further, this reservoir of gas is massive, containing up to three times more mass than gas within discs \citep{werk}. Even though the gas extends more than 150~kpc from the stellar disc, the metal enrichment implies that the gas originated within the stellar disc. What could loft so much mass out of discs? 

High-mass stars inject energy and momentum into the ISM through high energy photons, cosmic rays, and supernovae, together commonly called stellar feedback \citep{weaver, mckee77, chevalier, heckman90, murray05, thompson05, everett, socrates, hopkins12b, hopkins14, kim15, taeyson, bustard}. If the stellar feedback is highly concentrated, then the combined energy and momentum drives gas out of star forming regions into a galactic outflow \citep{heckman90, heckman2000, veilleux, erb15}. Galactic outflows may enrich the large reservoir of cirum-galactic gas, while also removing metals from low-mass galaxies to create the mass-metallicity relation \citep{tremonti04, finlator08, peeples11, andrews, zahid, creasey2015, christensen}. However, accurate mass outflow rates are required to constrain these relations.

Observations of galactic outflows are challenging. First, galactic outflows are multi-phase, with emission from hot X-ray emitting plasma \citep{griffiths, strickland2000, strickland09}, to ionized hydrogen \citep{lynds, bland88, shopbell, westmoquette2009a, newman, arribas}, and even cold molecular gas \citep{weiss99, matsushita2000, leroy15}. Second, outflows are diffuse, and probing outflows with emission lines is challenging and limited to only the local universe. Recent studies use optical \citep{heckman2000, martin2005, rupke2005b, chen10}, near UV \citep{weiner, martin09, steidel10, aleks, kornei12, erb12, rubin13} and far UV absorption lines \citep{pettini2000, pettini2002, shapley03, grimes09, claus2012, chisholm15, heckman2015, wood, chisholm16} to study the diffuse gas within outflows. While these studies have hinted at the role outflows play in removing gas from the discs of galaxies, the mass outflow rate calculations are plagued by uncertain assumptions.

To calculate the mass outflow rate from absorption lines, an observed column density and velocity are converted into a total mass outflow rate. The first step converts the observed ion density to a total Hydrogen density through an assumed metallicity and ionization correction \citep{rupke2005b, heckman2015}. The Hydrogen column density is then converted into a total Hydrogen mass through an assumed geometry, typically a thin shell. These studies make four key assumptions: (1) that the absorption lines are optically thin (2) that the measured ion is the dominant ionization state, and there is no ionization correction (3) that the metallicity of the outflow is consistent with the ISM of the galaxy and (4) that the outflow is a spherical shell with a radius of 5~kpc \citep{rupke2005b, weiner}. 

Recent observations question these assumptions. Ionization modelling from \citet{chisholm16} find that galactic outflows are photoionized, with the dominant ionization state, \siiiip, at an ionization potential near 25~eV, with only 1-5\% of the gas in the neutral phase. For neutral ions like Na~{\sc I}, neglecting the ionization correction can underestimate the total Hydrogen column density by a factor of 4 \citep{murray07, chisholm16}. Additionally, these photoionization models predict that the outflows are metal rich, and the assumption that the outflow has the metallicity of the ISM underpredicts the total Hydrogen density.

These large ionization fractions also imply that the outflow is relatively close to the ionizing source. Stars generate outflows at distances of super-star clusters, 20-40~pc \citep{mckee77, weaver, chevalier}, but metal-enriched gas is also observed out to 10~kpc from the starburst \citep{veilleux03, strickland04, rubin11, tumlinson}. Even though the outflow extends over this enormous distance, studies typically calculate the mass outflow rates as if the entire mass is in a thin shell with a radius of 5~kpc \citep{rupkee2005}.

Here, we present a new analysis of the mass outflow rate of the nearby galaxy NGC~6090 using {\it Hubble Space Telescope} ultraviolet spectroscopy. We calculate the mass outflow rate by first measuring the distance, metallicity, and ionization fraction of the outflow. In \autoref{data} we describe the data reduction, stellar continuum fitting, and measurement of the outflow properties. We then fit for how the optical depth and covering fraction scales with velocity, using a Sobelev optical depth, a $\beta$ velocity law, and a power-law density scaling (\autoref{beta}). We apply the ionization models of \citet{chisholm16} to the outflow of NGC~6090 to determine the metallcity (1.61~Z$_\odot$), total Hydrogen density (18.73~cm$^{-3}$), and the ionization fractions of the outflow (\autoref{ion}). In \autoref{ionstruct} we consider the implications of the ionization models and the metal enriched outflows, and then calculate the initial radius of the outflow (63~pc). Using these derived quantities, we examine the covering fraction (\autoref{cf}), velocity (\autoref{velocity}), and density (\autoref{density}) laws with radius. Since these relationships vary with velocity, we find that the mass outflow rate of the galaxy also varies with velocity, with a maximum mass outflow rate of 2.3~M$_\odot$~yr$^{-1}$. This mass outflow rate is 10 times smaller than it would be if we used previous assumptions, and only 9\% of the star formation rate of the galaxy. 

In this paper we use $\Omega_M = .28$, $\Omega_\Lambda = .72$ and H$_0$ = 70~\kms~Mpc$^{-1}$ \citep{wmap}.
\section{DATA}
\label{data}

\begin{figure}
\includegraphics[width = 0.5\textwidth]{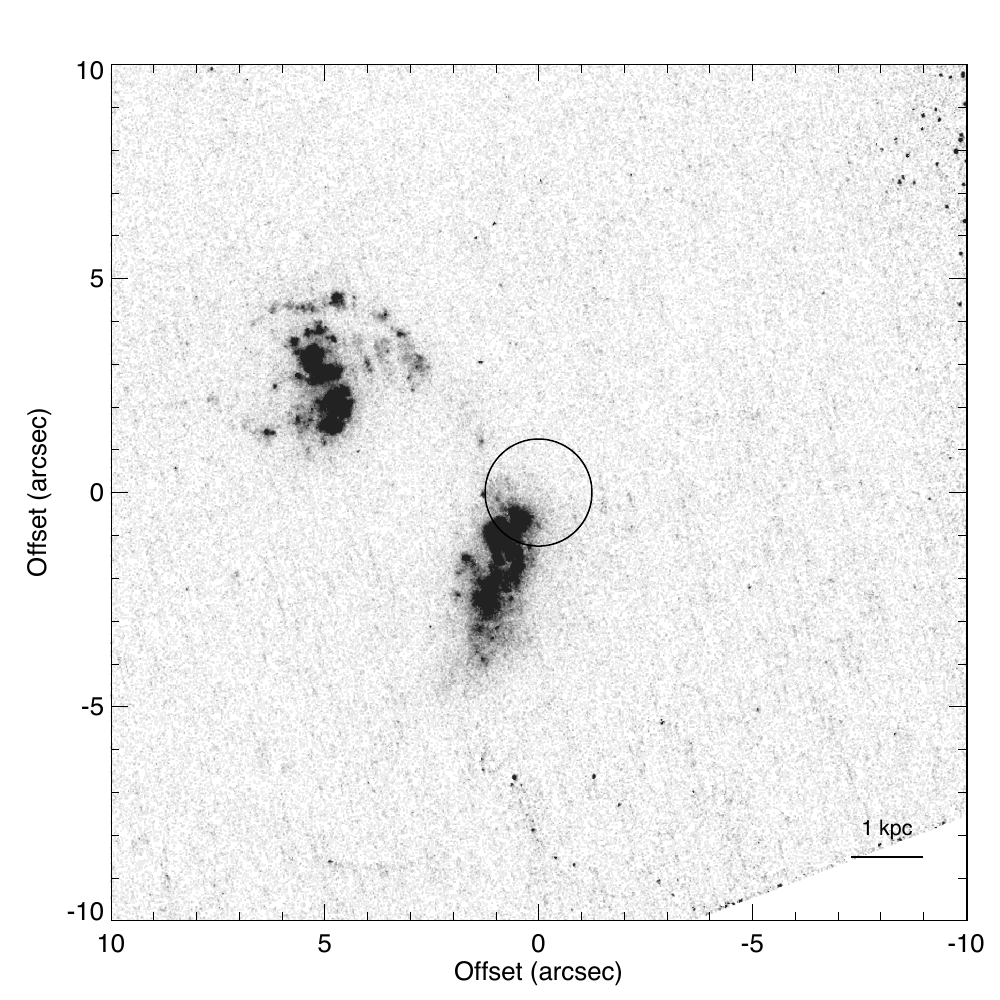}
\caption{Restframe UV image of NGC~6090 taken with the HST/ACS HRC camera and the F220W filter (restframe central wavelength of 2312~\AA). Two distinct nuclei are seen, along with a faint tidal-tail of UV emission in the bridge between the two nuclei. The approximate COS aperture location and size (2\farcs5 in diameter) are denoted by the black circle. A 1~kpc scale bar is given in the lower right corner, and 1\farcs0 is equal to 587~pc, at the distance of NGC~6090.}
\label{fig:image}
\end{figure}

\begin{table}
\begin{tabular}{ccc}
\hline
Row Number &Property &Value \\
\hline
(1) &log$_{10}($\mstarp/M$_\odot$) & 10.7 \\
(2) &SFR$_\mathrm{tot}$   &  25.15 M$_\odot$~yr$^{-1}$  \\
(3) &SFR$_\mathrm{COS}$   &  5.55 M$_\odot$~yr$^{-1}$  \\
(4) &Inclination & 29$^\circ$ \\
(5) &z & 0.0293 \\
(6) &D & 128~Mpc \\
(7) &log(O/H) + 12 & 8.77~dex (1.2~Z$_\odot$)  \\
(8) & log(SII/H$\alpha$) & -0.69 \\
(9) & log(OIII/H$\beta$) & -0.34 \\
(10) &E(B-V)  & 0.316 \\
(11) &Z$_\mathrm{s}$ &  1.0~Z$_\odot$\\
(12) &$<\mathrm{Age}>$ &   4.478~Myr \\
(13) & Resolution  &   48~\kms
\end{tabular}
\caption{Host galaxy properties of NGC~6090. The stellar mass (row 1), total star formation rate (SFR$_\mathrm{tot}$; row 2), extinction (row 10), stellar metallicity (row 11), and average stellar age (row 12) are calculated in \citet{chisholm16}, while the SFR within the COS aperture (SFR$_\mathrm{COS}$; row 3), the inclination (row 4), and the spectral resolution  (column 11) are from \citet{chisholm15}. The redshift (row 5) and distance (row 6) are taken from \citet{z}, while the gas phase metallicity (row 7) is from \citet{ismmetallicity}. The log(SII/H$\alpha$) and log(OIII/H$\beta$) values are from the JHU-MPA SDSS DR7 catalog \citep{sdss, brinchmann}, and show that the galaxy is within the star forming locus defined by \citet{kewley06}. }
\label{tab:ngc6090}
\end{table}
\begin{figure*}
\includegraphics[width = \textwidth]{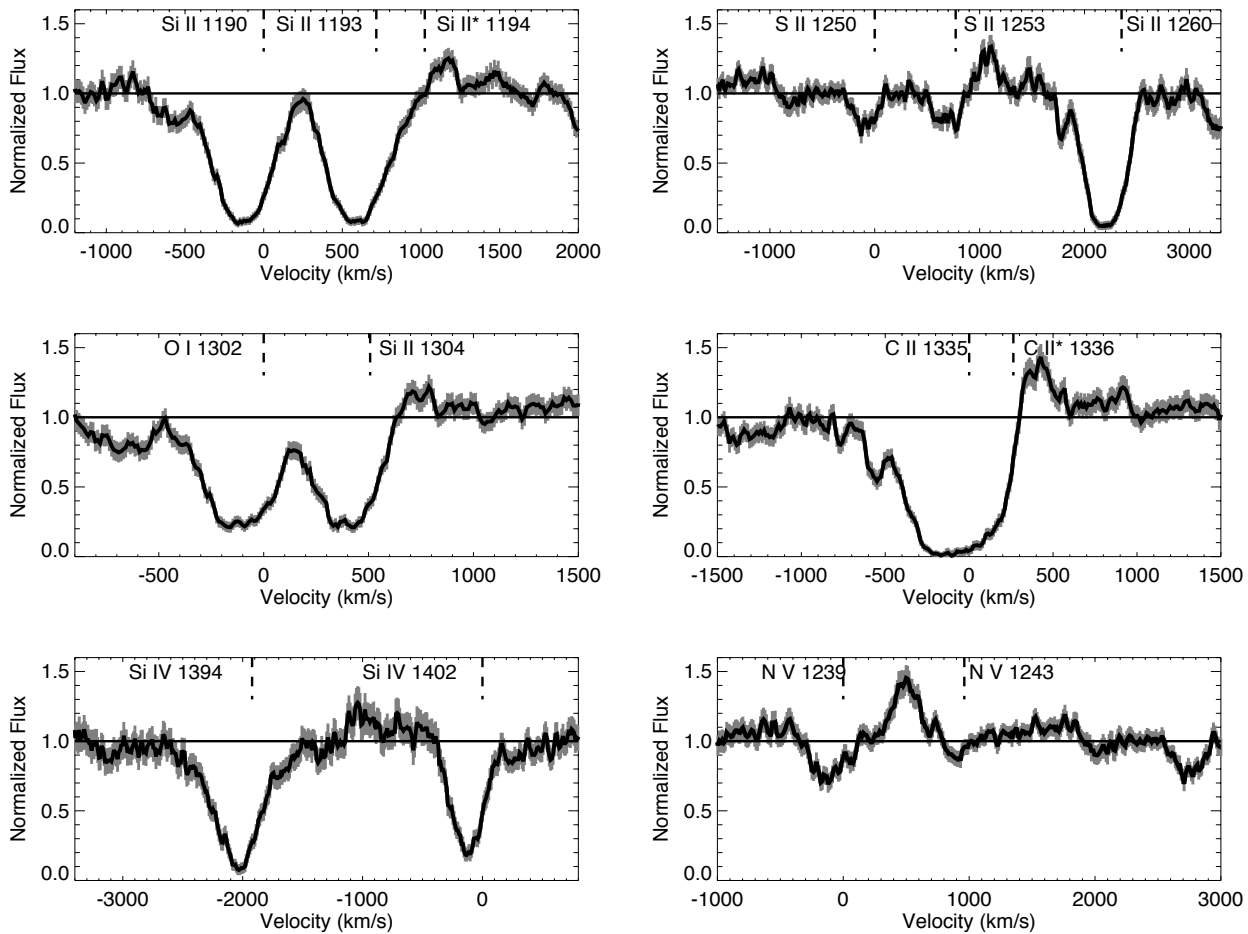}
\caption{Stellar continuum normalized absorption profiles as the thick black line. The gray envelope around the black line represents the 1$\sigma$ error on the flux measurement. Multiple transitions are plotted in each panel, and the velocity axis is only for the transition near zero velocity.  The zero velocities of all of the transitions are marked and labelled by vertical dashed lines in the upper portions of the plots.}
\label{fig:profile}
\end{figure*}

NGC~6090 is a massive star forming galaxy at a distance of 128~Mpc (see \autoref{tab:ngc6090}), with no indication of AGN contamination from the optical emission lines (see \autoref{tab:ngc6090}) and the {\it WISE} colors. The galaxy is metal rich, with a log(O/H)+12 of 8.77 \citep{ismmetallicity}. The galaxy is in the intermediate stages of a massive merger, as two distinct nuclei and large extend tidal tails are seen in the UV (\autoref{fig:image}) and optical images \citep{ostlin09}. 

The Cosmic Origins Spectrograph \citep[COS;][]{cos} on the {\it Hubble Space Telescope} observed NGC~6090 during Cycle 18 \citep[Project ID: 12173;][]{claus2012}. The full data reductions are given in \citet{chisholm15}, and here we summarize the steps taken to measure the optical depths, covering fractions, and column densities of the outflow. First, we describe the data reductions (\autoref{reduc}) and the stellar continuum fitting (\autoref{cont}). We then explore the effects of the Si~{\sc II} fine structure and resonance emission lines (\autoref{finestruct}). Finally, the optical depths, covering fractions, and column densities are measured in \autoref{measure}. 

\subsection{Data Reduction}
\label{reduc}

\citet{claus2012} observed NGC~6090  with the G130M grating for a total integration time of 8096~s. We downloaded the spectra from MAST and processed the data through the {\sc CalCOS} pipeline, version 2.20.1. The individual exposures were aligned and co-added following the methods outlined in \citet{wakker2015}. The spectrum was deredshifted using the redshift from \citet{z}. The flux was normalized to the median flux in a line free region between 1310-1320~\AA\ in the restframe, the wavelength array was then binned by 5 pixels (10~\kmsp at 1400\AA), and smoothed by 3 pixels. The set-up of the G130M grating affords continuous wavelength coverage between  1125-1425~\AA, in the restframe. Our fully processed data have a median signal-to-noise ratio of 12 per pixel after rebinning. 

\subsection{Stellar Continuum Fitting}
\label{cont}

Metal absorption in stellar atmospheres contaminates the ISM absorption profiles. To remove the stellar contribution, we fit the observed spectrum with a linear combination of {\sc STARBURST99} simple stellar populations models \citep{claus99, claus2010}. To match the resolution of the observations, and to avoid stellar libraries with Milky Way contamination, we use the fully theoretical Geneva models, with high-mass loss \citep{geneva94}, computed using the WM-basic method \citep{claus99, claus2010}.

A linear combination of different age stellar populations is required to simultaneously recover the \siiii photospheric lines near 1295~\AA\ (signatures of B stars) and the \siiv and \nv P-Cygni profiles (signatures of O stars). Since O and B stars dominate the FUV stellar continuum, we include {\sc STARBURST99} models of ten ages between 1-20~Myr, and use {\sc MPFIT} \citep{mpfit} to determine the linear combination of these ten models. Five ages contribute to the stellar continuum of NGC~6090 (1, 3, 4, 5, 20~Myr), but the 4 and 5~Myr models account for 84\% of the FUV light, for a weighted stellar age of 4.48~Myr (see \autoref{tab:ngc6090}). The stellar continuum fit is shown in the upper panel of Figure 2 in \citet{chisholm16}. This figure shows that the {\sc STARBURST99} models nicely reproduce the stellar P-Cygni features, and provide a continuum level for the spectrum.

Both the stellar age and metallicity determine the number of ionizing photons emitted by high-mass stars \citep{claus99}. Accordingly, we fit for both the metallicity and stellar age. {\sc STARBURST99} models have five stellar continuum metallicities -- 0.05, 0.2, 0.4, 1.0, 2.0~Z$_\odot$ -- and we determine that the stellar metallicity of NGC~6090 is solar using a $\chi^2$ test (\autoref{tab:ngc6090}). During the fitting, we include continuum reddening using a Calzetti extinction law \citep{calzetti}, and find that the stellar continuum is extincted by E(B-V) = 0.316~mags, after accounting for foreground Milky Way extinction \citep{schlegel}. We cross-correlate the fitted stellar continuum with the observed spectrum to establish the zero-velocity of the stellar continuum. Finally, we remove the stellar continuum by dividing the observed spectrum by the {\sc STARBURST99} model.

Strong Milky Way absorption and geocoronal emission lines are blueshifted from the metal lines of NGC~6090. However, the Milky Way lines are masked during the stellar population fitting to avoid contaminating the fit. Since the galaxy non-uniformly fills the COS aperture (see \autoref{fig:image}), the spectral resolution is degraded. The COS spectral resolution varies from 20~\kms for a point source to 200~\kms for a uniformly filled aperture \citep{france09}.  We fit the line width of the Milky Way absorption features, and find that the effective spectral resolution of the spectrum is 48~\kmsp. 

\autoref{fig:profile} gives the metal absorption profiles from the fully reduced, stellar continuum normalized spectra. There is a weak C~{\sc I}~1277\AA\ Milky Way absorption line near the \nv 1243~\AA\ transition and the stellar continuum is slightly underestimated. While we fit, and remove this weak C~{\sc I} feature, imperfect subtraction creates residual emission, contaminating the continuum near the weaker \nv 1243~\AA\ line. Therefore, we do not draw conclusions from the \nv profiles. 

\subsection{Fine Structure Emission Lines}
\label{finestruct}
%\begin{figure}
%\includegraphics[width = 0.5\textwidth]{si2star.pdf}
%\caption{Observed flux centered on the 1194\AA\ \siiifs emission line (black). A Gaussian fit to the line is shown in red. The \siii 1193\AA\ absorption feature is marked by the dashed vertical line, and the 1197\AA\ \siiifs line is  marked with the dot dashed line.}
%\label{fig:sifs}
%\end{figure}
Before we measure the outflow properties from the metal absorption lines, we must understand how resonance emission lines affect the absorption profiles \citep{prochaska2011, rubin11, scarlata, zhu15}. Continuum photons excite ground state electrons, and the electron then transitions into a lower energy state by emitting a photon with an energy of the difference between the two states. If the electron transitions back to the ground state through the same path as it was excited, it is called resonance emission. Resonant photons have the same restframe wavelength as the absorption and, depending on the geometry of the outflow, the resonance emission can overlap in velocity space with the absorption. This effectively reduces the measured absorption profile.

The amount, and the velocity, of the infilling can be estimated if fine structure splits the ground state. The electron's angular momentum splits the ground state into multiple levels, allowing the electron to transition through either resonance or fine structure emission (fine structure emission is normally denoted by a $^\ast$, i.e. as \siiifsp). The probability of transitioning by a specific path is given as $\mathrm{P}_\mathrm{i} = \mathrm{A}_\mathrm{i} / \Sigma \mathrm{A}_\mathrm{i}$, where A$_i$ is the Einstein A coefficient corresponding to level i, and the summation is over all possible levels. Therefore, we use the observed fine structure emission to estimate an upper limit on the in-filling of resonance profile.

There are four \siiifs lines in the observed wavelength regime at 1194, 1197, 1265, and 1309\AA, however the 1265 line is blended with the \oi geocoronal emission line. The 1194\AA\ line has the largest probability of fine structure emission (84\% of the 1190\AA\ absorption is re-emitted as 1194\AA\ \siiifsp), therefore this line predicts the maximum amount of  resonance emission. The 1194 \siiifs line is the strongest of the four \siiifs lines in the bandpass, but the equivalent width is only $-0.15$~\AA, a small fraction of the 1.74~\AA\ \siii 1190~\AA\ equivalent width. Further, the line is redshifted by +132~\kmsp, with emission only extending to -80~\kmsp. 

Below, we chiefly use the \siiv doublet, which does not have fine structure emission lines in the wavelength regime. The \siii and \siiv profiles have similar velocity profiles \citep{chisholm16}, allowing for the \siiifs to be used as a proxy for the \siiv emission. To avoid possible contributions from resonance emission, we exclude velocities greater than -80~\kms when fitting the profiles in \autoref{beta}.

\subsection{Equivalent Width, Optical Depth, Covering Fraction, and Column Density}
\label{measure}
\begin{table}
\begin{tabular}{cccc}
\hline
(1) & (2) &(3)  &(4) \\ 
Line & Equivalent Width & Velocity Range &log(N)  \\
 & (\AA) & (\kmsp)  &  (log(cm$^{-2}$)) \\
 \hline
\oi 1302 & 1.36 $\pm$ 0.02   & (-470, 150)  & $15.46 \pm 0.03$  \\
\siii 1190 & $1.74 \pm 0.03$ & (-730, 250) & $14.99 \pm 0.03$\\
\siii 1193 & $1.59 \pm 0.02$ & (-465, 300) & $14.64 \pm 0.02$\\
\siii 1260 &  $1.92 \pm 0.02$ & (-635, 205) & $14.40 \pm 0.02$  \\
\siii 1304 &  $1.12 \pm 0.02 $ & (-360, 190) & $15.12 \pm 0.03$  \\
\sii 1250  & 0.24 $\pm$ 0.02 & (-300, 110) & $15.50 \pm 0.17$ \\
\sii 1253 & 0.23 $\pm$ 0.02 & (-300, 110) & $15.19 \pm 0.18$  \\
\cii 1335 &   2.99 $\pm$ 0.02 & (-810, 300) & $15.59 \pm 0.01$ \\
\siiii 1206  &   2.58 $\pm$ 0.03 & (-780, 430) & $14.44 \pm 0.02$ \\
\siiv 1394  & 2.16 $\pm$ 0.04 & (-680, 430) & $14.64 \pm 0.04$\\
\siiv 1402  & 1.18 $\pm$ 0.03 & (-400, 160) & $14.63 \pm 0.04$ \\
\nv 1239  &  0.31 $\pm$ 0.02 &  ( -340, 130) & $14.22 \pm 0.16$\\
\nv 1243  &  0.07 $\pm$ 0.01 & (-170, 60) & $13.83 \pm 0.33$
\end{tabular}
\caption{Table of the absorption line properties for the 13 observed metal lines. The ions and wavelengths are given in column 1. The measured equivalent width, the velocity range integrated over, and the column density of each transition are given in columns 2, 3, and 4, respectively. Many of the strong lines suffer from saturation effects, therefore we only use column densities from the \siiv 1402\AA, \oi 1302\AA, \siii 1304\AA, and \sii 1250\AA\ lines. }
\label{tab:obs}
\end{table}
Once we have reduced the data, removed the stellar continuum, and explored the impact of the resonance emission features, we measure the properties of the outflow. Here, we describe the outflow with four parameters: equivalent width (W), optical depth ($\tau$), covering fraction (C$_f$), and column density (N).

W is measured by integrating $1-F_\mathrm{o}$ (the observed continuum normalized flux) over the velocity ranges given in column three of \autoref{tab:obs}. As discussed in \citet{chisholm16}, the maximum velocity of each transition strongly depends on the strength (the W) of the transition, not the ionization state of the transition. The W errors are computed by bootstrapping the observed flux with the flux error array 1000 times, and the standard deviation is calculated from the resultant W distributions. The equivalent width ratios of doublets (and triplets) can be used to diagnose the saturation of the transitions. With an observed doublet ratio of 1.83 (and an expected ratio of 2), the \siiv 1402\AA\ and 1392\AA\ lines are the only doublet that does not suffer from severe saturation effects. We also use the column densities of the weak \sii 1250\AA, \siii 1304\AA, and \oi 1302\AA\ lines during the ionization modelling (see \autoref{ion}).

%This is the calculation of the C_F and tau
C$_f$ is the fraction of the continuum source covered by the foreground absorbing gas (see \citet{rupkee2005} for a thorough description of the interpretation of C$_f$). At low $\tau$, C$_f$ and $\tau$ are degenerate because both impact the depth of the absorption profile. However, this degeneracy can be broken by solving the radiative transfer equation for the C$_f$ and optical depths of the lines \citep{hamann}. Here, we use the \siiv doublet because it does not suffer from strong saturation effects. The covering fraction is calculated in each pixel  to give a velocity (v) resolved C$_f$ profile as \citep{hamann}
\begin{equation}
C_f(\mathrm{v}) = \frac{\mathrm{F}_\mathrm{W}(\mathrm{v})^2-2\mathrm{F}_\mathrm{W}(\mathrm{v}) + 1}{\mathrm{F}_\mathrm{S}(\mathrm{v})-2\mathrm{F}_\mathrm{W}(\mathrm{v})+1}
\label{eq:cf}
\end{equation}
where F$_\mathrm{W}$ is the continuum normalized flux of the weaker doublet line (\siiv 1402\AA), and F$_\mathrm{S}$ is the continuum normalized flux of the stronger doublet line (\siiv 1394). In this equation we use the fact that the \siiv 1402 line has half the oscillator strength of the 1392 line. We set unphysical C$_f$ values that are less than zero to zero. We also calculate the velocity resolved $\tau$ as
\begin{equation}
\tau(\mathrm{v}) = \ln\left(\frac{C_f(\mathrm{v})}{C_f(\mathrm{v})+F_\mathrm{W}(\mathrm{v})-1}\right)
\label{eq:tau}
\end{equation}
We bootstrap the errors of C$_f$ and $\tau$, similar to the W errors above.

Finally, we calculate the integrated column density of each transition using the apparent optical depth method \citep{savage}. Since some of the weak lines are singlets, we cannot use \autoref{eq:cf} to calculate the column density, rather we have to assume that the lines are fully covered. Below, we find that this assumption is fair for the line centers, which have most of the column density. The integrated column density is calculated as
\begin{equation}
    \mathrm{N} = \frac{3.77 \times 10^{14}~\text{cm}^{-2}}{\lambda[\text{\AA}] f} \int \text{ln} 
    \frac{1}{F_\mathrm{o}(\mathrm{v})} \mathrm{dv}
\end{equation}
where f is the oscillator strength \citep{chisholm16}, and the total N is calculated by integrating over the velocity interval given in \autoref{tab:obs}. With the measured properties of the outflow, we now study how the \siiv 1402\AA\ optical depth and covering fraction evolve with velocity.

\section{PROFILE FITTING}
\label{beta}

\begin{figure*}
\includegraphics[width = \textwidth]{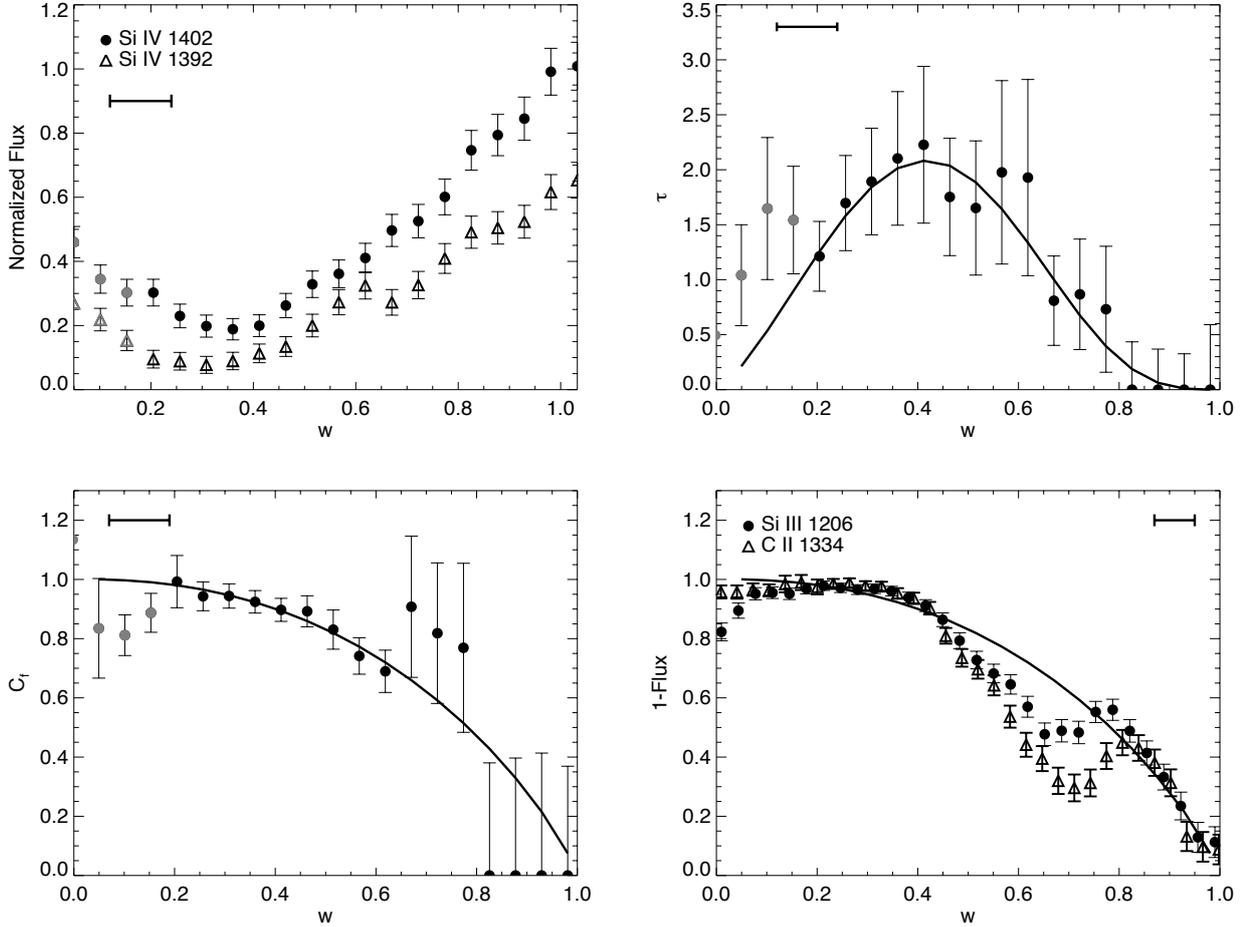}
\caption{Normalized velocity ($w = \mathrm{v}/\mathrm{v}_\infty$; $\mathrm{v}_\infty =$-399~\kmsp) profiles of the \siiv 1402~\AA\ absorption line. The upper left panel shows the absorption profiles of the \siiv 1402~\AA\ (solid circles) and \siiv 1392~\AA\ lines (open triangles). The upper right panel shows the variation of the \siiv 1402~\AA\ optical depth ($\tau$) with normalized velocity computed using the radiative transfer equation for the \siiv doublet (\autoref{eq:tau}). The lower panel shows the variation of the \siiv covering fraction (C$_f$; \autoref{eq:cf}) with normalized velocity. The black lines show the simultaneous best fit of \autoref{eq:cfbeta} for $\tau$ and C$_f$. Velocities less than 0.2 are excluded from the fits due to contributions from resonant emission and zero-velocity absorption (\autoref{finestruct}), and are plotted in gray. The best-fit parameters are given in \autoref{tab:beta}. Lower right panel: velocity profile of 1-Flux, an approximation for C$_f$ for optically thick lines, for the \siiii (circles) and \cii (triangles) profiles. These two lines are amongst the strongest metal lines in the bandpass. The \siiv C$_f$ fit from the lower left panel is over-plotted, demonstrating that the \siiv C$_f$ fit is also a fair representation of the saturated lines.  The velocity resolution (48~\kmsp) is given by a bar in the corner of each panel.}
\label{fig:beta}
\end{figure*}

In \autoref{fig:profile} the metal absorption lines are broad ($\sim175$~\kmsp) and blueshifted by $\sim-150$~\kmsp. Under the Sobolev approximation, these profiles are produced by an expanding medium, that has a steep velocity gradient (dv/dr) but a narrow intrinsic line profile, such that the expanding medium broadens the line profile by absorbing and scattering continuum photons \citep{sobolev, lamers, prochaska2011, scarlata}. The Sobolev optical depth of the \siiv 1402\AA\ profile is given by
\begin{equation}
\tau\left(\mathrm{v}\right) = \frac{\pi e^2}{\mathrm{mc}} f \lambda~\mathrm{n}_\mathrm{4}\left(\mathrm{r}\right) \frac{\mathrm{dr}}{\mathrm{dv}}
\label{eq:sobelv}
\end{equation} 
where $f$ is the oscillator strength (0.255 for \siivp), $\lambda$ is the rest wavelength, and n$_\mathrm{4}$ is the number density of the \siiv gas at a given radius (r). To simplify the calculation, we introduce normalized quantities by dividing the velocity by the maximum velocity ($w = \mathrm{v}/\mathrm{v}_\infty$, v$_\infty = -399$~\kms for \siiv 1402\AA) and the radius by the inner radius ($x = \mathrm{r}/\mathrm{R}_\mathrm{i}$), which we measure in \autoref{ionstruct}. v$_\infty$ is calculated as the velocity at which the absorption reaches 90\% of the continuum, similar to the method outlined in \citet{chisholm15}.

We must model how n$_\mathrm{4}$ is distributed with radius. Previous studies assume that the density is either constant with velocity \citep{martin09, steidel10}, or that the density follows the continuity equation \citep{prochaska2011, scarlata}. However, to remain general we assume that the density follows a power-law scaling with radius, such that
\begin{equation}
\mathrm{n}_\mathrm{4}(x) = \mathrm{n}_\mathrm{4,0} x^\alpha   
\label{eq:den}
\end{equation}
where n$_\mathrm{4,0}$ is the \siiv density at the initial radius, R$_\mathrm{i}$. Placing this into \autoref{eq:sobelv} gives the relation
\begin{equation}
\tau (x) = \frac{\pi e^2}{\mathrm{mc}} f \lambda_\mathrm{0} \frac{\mathrm{R}_\mathrm{i}}{\mathrm{v}_\infty} \mathrm{n}_\mathrm{4,0} x^\alpha \frac{dx}{dw} = \tau_0 x^\alpha \frac{\mathrm{d}x}{\mathrm{d}w}
\label{eq:sobelvnorm}
\end{equation}
where we have combined the constants into a single constant, $\tau_0$, that represents the maximum optical depth.

Additionally, we derive the covering fraction of the \siiv absorption (C$_f$; see \autoref{eq:cf}). As we discuss in \autoref{cf}, the covering fraction is the proportion, expressed as a percentage, of the starburst area covered by the outflow at a given radius. We assume that the covering fraction scales as a power-law with distance (as physically motivated in \autoref{cf}), such that
\begin{equation}
C_f(w) = C_{f} (\mathrm{R}_\mathrm{i}) \left(\frac{\mathrm{r}}{\mathrm{R}_\mathrm{i}}\right)^\gamma = C_{f} (\mathrm{R}_\mathrm{i}) x^\gamma
\label{eq:cffit}
\end{equation}
Now we have related the two observables of the profile to the distance from the starburst.

\autoref{eq:sobelvnorm} and \autoref{eq:cffit} are given in terms of distance, while we measure the parameters in terms of velocity. Many of the physical mechanisms for driving outflows scale the velocity with radius as a $\beta$ velocity law \citep[see \autoref{velocity} below;][]{lamers}, such that
\begin{equation}
\mathrm{v} = \mathrm{v}_\infty \left(1-\frac{\mathrm{R}_\mathrm{i}}{\mathrm{r}}\right)^{\beta} 
\label{eq:beta}
\end{equation}
Substituting for w and x, and inverting this relation gives the normalized radius in terms of the velocity as
\begin{equation}
x = \frac{1}{1-w^{1/\beta}}
\end{equation}
Taking the derivative of this gives the velocity gradient in terms of the normalized velocity and the $\beta$ exponent as
\begin{equation}
\frac{\mathrm{d}x}{\mathrm{d}w} = \frac{w^{1/\beta-1}}{\beta\left(1-w^{1/\beta}\right)^2}
\end{equation}
We can also use the $\beta$-law to derive the density scaling with velocity as 
\begin{equation}
\mathrm{n}(w) = \mathrm{n}_\mathrm{4,0} \left(\frac{1}{(1-w^{1/\beta})}\right)^{\alpha}
\label{eq:continuity}
\end{equation}
producing an $\tau$ (\autoref{eq:sobelvnorm}) and C$_f$ (\autoref{eq:cffit}) velocity scaling as 
\begin{equation}
\begin{aligned}
\tau\left(w\right) &= \tau_\mathrm{0} \frac{w^{1/\beta-1}}{\beta(1-w^{1/\beta})^{2+\alpha}}\\
C_f(w) &=  \frac{C_f (\mathrm{R}_\mathrm{i})}{\left(1-w^{1/\beta}\right)^\gamma}
\label{eq:cfbeta}
\end{aligned}
\end{equation}
We then have five parameters to fit for: the maximum optical depth ($\tau_0$), the covering fraction at the initial radius (C$_{f}(\mathrm{R}_\mathrm{i})$), the exponent of the beta velocity-law ($\beta$), the exponent of the density law ($\alpha$), and the exponent of the covering fraction law ($\gamma$). 

Using {\sc MPFIT} \citep{mpfit}, we simultaneously fit for the five parameters from the unsaturated \siiv 1402~\AA\ $\tau$ and C$_f$ distributions (see \autoref{fig:beta}). At w below 0.2 ($v > -80$~\kmsp) zero-velocity absorption and resonance emission may effect the distributions (see \autoref{finestruct}); therefore, we only fit the distributions between w of 0.2 and 1.0 (the points outside this range are coloured gray in \autoref{fig:beta}). The C$_f$ and $\tau$ fits are shown in \autoref{fig:beta}, and the fit parameters are given in \autoref{tab:beta}. In these fits we have binned the flux array by a factor of two (20~\kmsp) to increase the signal-to-noise ratio, while still Nyquist sampling the observed velocity resolution (see the bar in \autoref{fig:beta}). These fits reproduce the observed $\tau$ and C$_f$ distributions, and describe the radial velocity and density laws of the outflow through \autoref{eq:beta} and \autoref{eq:continuity}. In \autoref{discussion} we discuss some implications of these trends.

To test the covering fraction fits, we can use the \siiii and \cii profiles. Since these profiles are much stronger, they are strongly saturated at all velocities. This means that we cannot measure the C$_f$ with the previous formula (\autoref{eq:cf}), while $\tau$ cannot be accurately measured at all. Rather, at large $\tau$, the depth of the line is completely set by C$_f$, and is measured as C$_f(\mathrm{v}) = 1 - F(\mathrm{v})$. In the lower right panel of \autoref{fig:beta} we plot the C$_f$ for \siiii (circles) and \cii (triangles). The  \siiv C$_f$ fit is over-plotted on these measurements and provides reasonable agreement with the data.

\begin{table}
\begin{tabular}{cccccc}
\hline
(1) & (2) & (3) & (4) & (5) \\
$\tau_0$ & $\beta$ & $\alpha$ & C$_{f} (\mathrm{R}_\mathrm{i})$ & $\gamma$\\
\hline
 $4.80\pm 1.41$ & $0.43\pm 0.07$ &  $-5.72 \pm 1.51$ & $1.00 \pm 0.04$ & $-0.82\pm 0.23$
\end{tabular}
\caption{Table of the quantities derived from the profile fits of the optical depth ($\tau$) and covering fraction (C$_f$; see \autoref{eq:cfbeta}). The individual parameters are: Column 1, $\tau_0$, optical depth at line center; Column 2, $\beta$, the velocity power-law index; Column 3, $\alpha$, the column density power-law index; Column 4, C$_f (\mathrm{R}_\mathrm{i})$, the maximum covering fraction; Column 5, $\gamma$, the covering fraction power-law index.}
\label{tab:beta}
\end{table}

\section{IONIZATION MODELLING}
\label{ion}

In \autoref{measure} we calculate the integrated column density (N) for individual metal ions, but these only describe the individual ion and not the total mass of the outflow. While Ly-$\alpha$ absorption traces the neutral Hydrogen in the outflow, we opt to use the metal lines to describe the outflow because (1) the strong P-Cygni Ly-$\alpha$ profile requires detailed radiative transfer models to constrain \citep{verhamme} and (2) up to 99\% of the Hydrogen in the outflow may be ionized \citep{chisholm16}. To calculate the total Hydrogen in the outflow, ionization models are required to measure the ionization fraction ($\chi_\mathrm{i}$, or the fraction of the total gas within transition i) and the abundance (N$_\mathrm{i}$/N$_\mathrm{H}$, or the ratio of an element to the total Hydrogen).

In \citet{chisholm16} we find that photoionization models describe the ionization structure of galactic outflows. These models require the outflow metallicity  (Z$_\mathrm{o}$) to be greater than 0.5~Z$_\odot$ and the ionization parameter (log(U); the ratio of the photon density to outflow density) to be between -1.5 and -2.25. Additionally, the ionization structure depends both on the spectral energy distribution and the strength of the ionizing source, as well as the density (n$_\mathrm{0}$) of the outflow. 

We model the ionization structure using {\sc CLOUDY} version 13.03 \citep{ferland}. The {\sc CLOUDY} models are not velocity resolved profiles, rather we are fitting the measured integrated column densities to the integrated values from {\sc CLOUDY}. We assume that the outflow is ionized by the observed stellar continuum, using the best-fit {\sc STRABURST99} stellar continuum model from \autoref{cont}. These {\sc STARBURST99} models have an age of 4.478~Myr and a constant SFR of 25.15~M$_\odot$~yr$^{-1}$ (see \autoref{tab:ngc6090}). In the {\sc CLOUDY} models, we assume an expanding spherical geometry, which is not meant to reproduce the velocity profiles, but to account for back-scattering of radiation. We set the covering fraction to 1.00, the observed value from the \siiv transition, and scale the density with a power-law of -5.72, as measured above. The {\sc CLOUDY} models are stopped once the simulations reach 3000~K, which is lower than the default criteria to allow for higher metallicities. These lower temperatures require a cosmic ray background to be included \citep{indriolo}.

We use {\sc CLOUDY}'s default H~{\sc II} abundances, which are similar to the Milky Way values \citep{baldwin1991, savage, osterbrock1992, rubin1993}, and include an Orion Nebular dust grain distribution \citep{baldwin1991}. The dust grains account for scattering and destruction of photons, as well as depletion of metals onto grains. The abundances are scaled by a constant factor to change the outflow metallicity (Z$_\mathrm{o}$). Similarly, we vary the outflow density at the inner radius (n$_\mathrm{0}$), ionization parameter (U), and the stellar continuum metallicity (Z$_\mathrm{s}$). 

We use a Bayesian approach to estimate Z$_\mathrm{o}$, n$_\mathrm{0}$, U, and Z$_\mathrm{s}$ \citep{kauffmann, brinchmann, stats}. While  \citet{chisholm16} uses the W ratios to define the ionization structure, the relatively high signal-to-noise ratio and spectral resolution observations of NGC~6090 allow us to use the measured column densities from four weak transitions: \oi 1302\AA, \siii 1304\AA, \sii 1250\AA, and \siiv 1402\AA\ (see \autoref{tab:obs}). We do not use the \nv column densities due to Milky Way C~{\sc I}1277\AA\ and imperfect continuum subtraction contamination near the \nv line (see \autoref{cont}). We create grids of {\sc CLOUDY} models, with differing input parameters (\autoref{tab:gridmod}), and tabulate the predicted {\sc CLOUDY} column densities. We assume a uniform prior -- each model is equally likely -- and compute the likelihood function of each set of parameters as
\begin{equation}
\mathrm{L} \propto \exp(-\chi^2/2).
\end{equation}
where $\chi^2$ is the chi-squared function using the observed column density (N), the errors on the observed N, and the predicted N from {\sc CLOUDY}. For each parameter (Z$_\mathrm{o}$, n$_\mathrm{0}$, U, and Z$_\mathrm{s}$), the likelihood function is marginalized over the other nuisance parameters and normalized to one to create probability distribution functions (PDFs). We calculate expectation values and standard deviations of the individual parameters from these PDFs as estimates of the parameters, and their uncertainties.
\begin{figure}
\includegraphics[width = 0.5\textwidth]{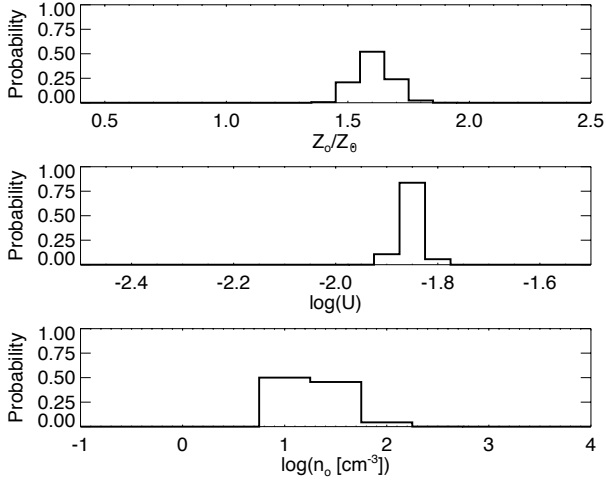}
\caption{Probability density functions (PDFs) from the {\sc CLOUDY} ionization modelling. The three parameters are outflow metallicity (top panel), ionization parameter (middle panel), and the total Hydrogen density at the initial radius (bottom panel). The PDFs are single peaked, with expectation values and standard deviations given in \autoref{tab:ion}.}
\label{fig:ionpdfs}
\end{figure}

We do the fitting in two iterations. The first iteration uses a course grid to determine the best Z$_s$. Since there are only five stellar continuum metallicities in the fully theoretical {\sc STARBURST99} models, we only coarsely estimate the stellar continuum metallicity. Similar to the COS spectrum fitting (see \autoref{cont}), the Bayesian analysis finds that the 1~Z$_\odot$ model best fits the ionization structure, assigning 100\% of the probability to the 1~Z$_\odot$ stellar continuum model.

\begin{table}
\begin{tabular}{cccc}
\hline
Parameter & Range & Step Size & Number of Steps \\
\hline
log(U) & (-2.5, -1.5) & 0.05 & 21  \\
Z$_\mathrm{o}$/Z$_\odot$ & (0.4, 2.5) & 0.1 & 22 \\
log(n$_\mathrm{0}$) & (-1, 4) & 0.5 & 11  \\
\end{tabular}
\caption{Grid of {\sc CLOUDY} models used in the Bayesian analysis of the ionization structure.}
\label{tab:gridmod}
\end{table}
The second iteration uses a more finely spaced grid for Z$_o$, log(U) and n$_0$, while only using the solar metallicity stellar continuum model (see \autoref{tab:gridmod}). In \autoref{fig:ionpdfs} we show the PDFs for the three parameters, with narrow distributions that have expectation values of $Z_\mathrm{o} = 1.61\pm0.08$~Z$_\odot$, $n_\mathrm{0} = 18.73\pm2.37$~cm$^{-3}$, and  log(U)$ =-1.85\pm0.02$ (see \autoref{tab:ion}). Using these parameters, we create a best-fit {\sc CLOUDY} model, producing the ionization fractions of the outflow (see \autoref{tab:ion}). As discussed in \autoref{ionstruct}, these ionization models have important implications for the metallicity of the outflow, the inner radius of the outflow, and the mass outflow rate.

\begin{figure}
\includegraphics[width = 0.5\textwidth]{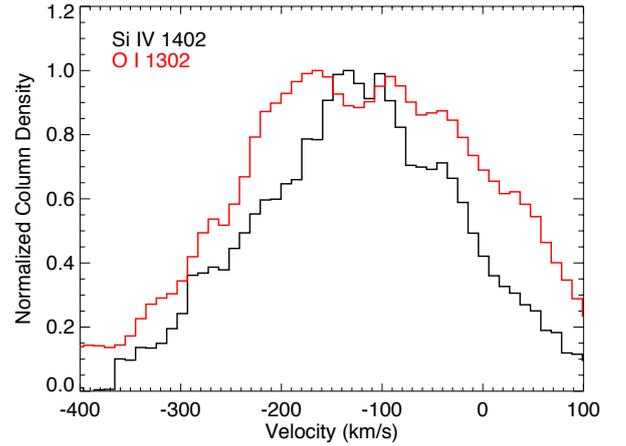}
\caption{The velocity resolved column density plots for the \siiv (black) and \oi (red) lines, normalized to their maximum values. The velocity structure between the neutral (\oip) and highly ionized gas (\siivp) stays roughly constant over the velocity range, implying that the ionization structure does not vary substantially with velocity.}
\label{fig:colfrac}
\end{figure}

\begin{table}
\begin{tabular}{ccc}
\hline
Row Number  & Property & Value \\
\hline
(1) & log(U) & -1.85 $\pm$ 0.02 \\
(2) & Z$_\mathrm{o}$/Z$_\odot$ & 1.61 $\pm$ 0.08 \\
(3)& Z$_\mathrm{s}$/Z$_\odot$ & 1.0  \\
(4) & n$_\mathrm{0}$ [cm$^{-3}$] & 18.73 $\pm$ 2.37  \\
(5) & $\chi_\mathrm{Si~{\sc IV}}$ & $0.087^{\scriptsize +0.004}_{\scriptsize -0.002}$  \\
(6) & log(Si/H) & -5.19$\pm0.02$ \\
(7) &log(N$_\mathrm{Si~{\sc IV}}$) [log(cm$^{-2}$)] & 14.64 $\pm$ 0.04 \\
(8) & log(N$_\mathrm{H}$) [log(cm$^{-2}$)] & 20.89$^{\scriptsize +0.04}_{\scriptsize -0.03}$ \\
(9) & R$_\mathrm{i}$ [pc] & 63.4  \\
(10) & T$_\mathrm{H}$ [K] & 5470 \\
\end{tabular}
\caption{Table of the quantities derived from the {\sc CLOUDY} photoionization modelling. Row (1) is the ionization parameter; (2) is the derived outflow metallicity; (3) is the stellar continuum metallicity; (4) is the Hydrogen density at the base of the outflow; (5) is the ionization fraction of \siivp, or the per cent of the total Si in the \siiv transition; (6) The best-fit Si to H abundance value; (7) The measured \siiv column density; (8) The total Hydrogen column density calculated from the \siiv column density, Si/H abundance, and the \siiv ionization fraction; (9) The inner radius, calculated using \autoref{eq:r}; (10) the H~{\sc II} temperature of the best-fit {\sc CLOUDY} model. The errors on $\chi_\mathrm{Si {\sc IV}}$ and log(Si/H) are calculated by producing {\sc CLOUDY} models of the estimate plus/minus 1$\sigma$ errors of the Z$_\mathrm{o}$, n$_0$, and log(U).}
\label{tab:ion}
\end{table}

\subsection{Ionization Structure With Velocity}
\label{ionvel}
In the following analysis we use an integrated ionization correction. However, if the outflow is heated as it is accelerated, then the column density ratio of high to low ionization potential lines should increase with velocity. \autoref{fig:colfrac} shows the velocity resolved column densities  for the \siiv and \oi transitions, showing that there is not a coherent variation of the column densities with velocity between $-400$ and 0~\kmsp. This implies that a majority of the column density arises from a small radius, and the photoionization models are approximately plane-parallel.   

\section{DISCUSSION}
\label{discussion}
Here we discuss various aspects of the profile and ionization fitting from \autoref{beta} and \autoref{ion}. In \autoref{ionstruct}, we first derive important parameters of the outflow. We then study the implications for the covering fraction (\autoref{cf}), velocity (\autoref{velocity}), and density (\autoref{density}) radial scalings. Finally, we combine the various relations to determine the mass outflow rate of NGC~6090 (\autoref{mout}).

\subsection{Ionization Modelling}
\label{ionstruct}

With the ionization fractions and metallicities derived in \autoref{ion}, we convert the measured \siiv column densities into a total Hydrogen density (see \autoref{tab:ion}). The total (neutral plus ionized) Hydrogen column density is 20.89$_{-0.03}^{+0.04}$~cm$^{-2}$, where the uncertainties for each parameter are propagated to compute the uncertainties of the total Hydrogen column density. The UV continuum extinction provides a complimentary way to estimate the total Hydrogen column density by relating the extinction to the amount of dust through a dust-to-gas ratio. We use the relation from \citet{claus2002} and \citet{heckman2011} to calculate the total Hydrogen column density as
\begin{equation}
    \mathrm{N}_\mathrm{H} = \frac{3.6 \times 10^{21} \mathrm{E(B-V)}}{Z} \text{cm}^{-2}
\end{equation}
where we use a starburst dust attenuation law and the relation for the FUV optical depth from \citet{gildepaz} to convert the spectral slope to E(B-V). Using the extinction measured from the UV continuum in \citet{chisholm16} and the outflow metallicity from the ionization structure (see \autoref{tab:ngc6090} and \autoref{tab:ion}), we expect a log(N$_\mathrm{H}$) of 20.85, in agreement with the derived total N$_\mathrm{H}$ from the ionization modelling.

The ionization model predicts that 99.3\% of the Hydrogen in the outflow is ionized. These low neutral fractions suggest that the outflow has a H~{\sc I} column density of $5 \times 10^{18}$~cm$^{-2}$. Current radio arrays cannot detect this H~{\sc I} column in emission, but the upcoming SKA will detect these densities with 10\farcs0 spatial resolution and 5~\kms spectral resolution in 10-100 hours of integration time \citep{ska}. 

The outflow metallicity is 1.61~Z$_\odot$, 61\% and 34\% larger than the measured stellar and ISM metallicities (see \autoref{tab:ngc6090}). Substantial uncertainties on log(O/H) measurements \citep{kewley06} means that the outflow metallicity is at least consistent with the ISM values, and likely enriched compared to the ISM of the galaxy. Additional metals may reside in a hotter phase not probed by these observations (see below for a discussion of this phase). Metal enriched outflows are important constraints for models using galactic outflows to explain the mass-metallicity relation \citep{tremonti04, finlator08, peeples11, creasey2015, christensen}. For example, to match the mass-metallicity relation of \citet{denicolo} for galaxies with log(\mstarp/M$_\odot$) of 10.7, Figure 8 of \citet{peeples11} suggests that the outflow metallicity is 1.58~Z$_\odot$. The same figure also uses Z$_\mathrm{o}$ to constrain the scaling of the mass outflow rate divided by the star formation rate (mass-loading factor). In future studies, we will determine the scaling of the outflow metallicity with stellar mass and star formation rate to better constrain these studies.

The initial radius (R$_\mathrm{i}$) can be determined from the fitted \siiv optical depth (see \autoref{eq:sobelvnorm}). Solving for R$_\mathrm{i}$ in terms of the observed maximum \siiv optical depth, we find that
\begin{equation}
    \mathrm{R}_\mathrm{i} = \frac{\mathrm{mc}}{\pi e^2} \frac{1}{f \lambda} \frac{\mathrm{v}_\infty \tau_\mathrm{0}}{\mathrm{n}_\mathrm{0} \chi_\mathrm{Si~IV} \mathrm{Si/H}} = 63.4~\text{pc}
    \label{eq:r}
\end{equation}
Using the values from the $\beta$-profile fitting (\autoref{tab:beta}) and the ionization modelling (\autoref{tab:ion}). We have related the initial \siiv density ($n_{4,0}$) to the Hydrogen density at the base of the outflow, the abundance, and the ionization corrections from the ionization models ($n_{4,0} = n_\mathrm{0} \chi_\mathrm{Si IV} \mathrm{Si/H}$). The inner outflow radius and the size of the mass-loading region (maximum extent of about 150~pc; see \autoref{mout} and \autoref{fig:mout} below) is consistent with models of the mass-loading region in M~82 \citep{suchkov, strickland09}, and with analytical work from \citet{heckman2015}, who assume that the inner outflow radius is twice the size of the starburst. However, what is the physical meaning of the inner outflow radius?

The shredding of supernovae blastwaves by hydrodynamical instabilities is a possible origin for this inner radius. Supernovae energy builds up in the ISM due to multiple impulsive events \citep{sharp}, which creates a blastwave that shocks when it encounters the ambient ISM \citep{taylor, sedov, weaver, mckee77, mckee88, kim15}. The shock compresses surrounding ISM into a thin, dense shell of relatively cool gas that travels outward. This phase is called the snowplow phase, and the radius depends on the amount of injected energy, the density of the medium, and how much of the energy is radiated away \citep{weaver, mckee77, cooper08, sharp, kim15}, with typical values near 20-30~pc \citep{draine, kim15}.

Once the shell forms, Rayleigh-Taylor and Kelvin-Helmholtz instabilities rapidly destroy it, creating many small warm (10$^4$~K) cloudlets \citep{maclow, cooper08, fujita09, martin09, sharp, thompson16}. The destruction of the blastwave allows the hot interior gas to escape and travel into the halo as a hot wind. The hot, high-velocity wind, may then accelerate these clouds through ram pressure to the observed velocities \citep{fujita09, martin09, thompson16}. Therefore, a possible origin for R$_\mathrm{i}$ is that hydrodynamic instabilities have shredded the dense blastwave, and injected cloudlets into the hot wind. 

Finally, the {\sc CLOUDY} modelling estimates the temperature of the H~{\sc II} in the outflow to be 5470~K (see \autoref{tab:ion}), leading to a pressure of P/k$_\mathrm{b}$ = $1 \times 10^5$~K~cm$^{-3}$. The hot wind is typically assumed to have a temperature of 10$^7$~K \citep{chevalier}. If the observed cloudlets and hot wind are in pressure equilibrium, then the hot wind has a density of 0.01~cm$^{-3}$ at R$_\mathrm{i}$, a factor of 10 lower than the density in the \citet{chevalier} model. However, recent models by \citet{bustard} show that the density and temperature of this hot wind can greatly vary, depending on the efficiency and mass-loading of the outflow. 

Regardless of the origin of the outflow, it must be evenly distributed across the stellar continuum to produce a unity covering fraction (see \autoref{tab:beta}). However, the external pressure on the outflow decreases as it accelerates out from the starburst. The change in pressure has a dramatic effect on the size of the cloudlets in the outflow, and in turn their covering fraction. In the next section we discuss this pressure change, and how it naturally leads to the observed scaling of the covering fraction with velocity.

\subsection{Covering Fraction}
\label{cf}
\begin{figure}
\includegraphics[width = 0.5\textwidth]{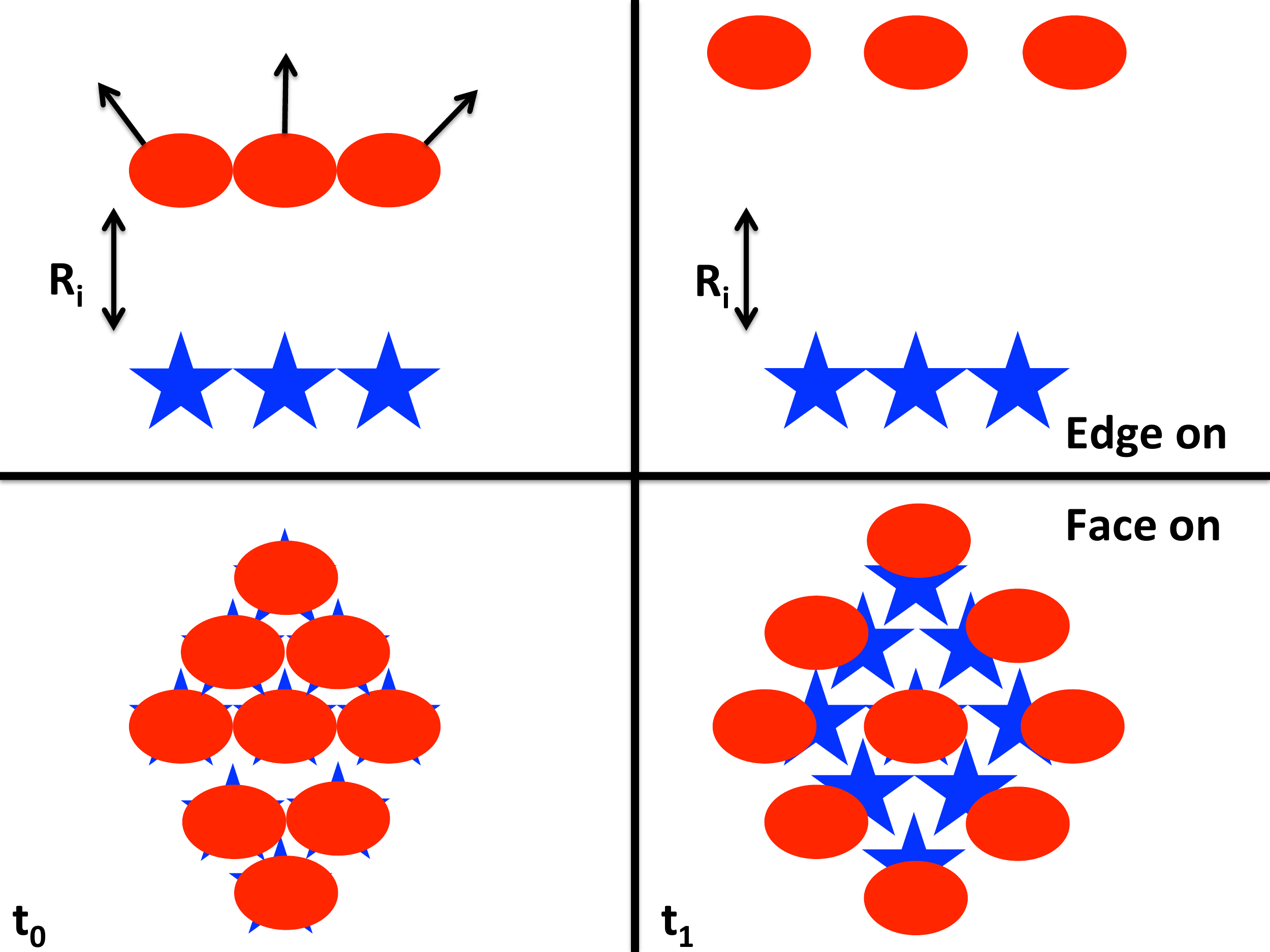}
\caption{Upper panels: Edge on cartoon of outflowing cloudlets (red circles).  At an initial time, t$_0$, (left panels) the clouds start a distance, R$_\mathrm{i}$, from the starburst (blue stars). At t$_0$, these clouds completely cover the background stars (see face on view in the lower left panel), and the stellar continuum is not transmitted. The clouds are accelerated away from the starburst until a later time, t$_1$ (right panels). We assume that these clouds maintain their size, and gaps are created between the cloudlets as they move away from the starburst. When viewed face on (lower right), the gaps expose the background starburst, reducing the covering fraction as x$^{-2}$ (see text).}
\label{fig:cfnop}
\end{figure}

In \autoref{beta}, we observe that the outflow initially completely covers the background stellar continuum, but as the velocity increases the outflow covers less of the background stars (C$_f$ drops as $x^{-0.8}$). To physically understand this scaling relation, we hypothesize that the absorption arises from cloudlets of gas initially a distance R$_\mathrm{i}$ from the starburst (see the upper left panel of \autoref{fig:cfnop}). At R$_\mathrm{i}$ these clouds are large enough to completely occupy the volume along the line-of-sight to the stars: none of the stellar continuum is transmitted (see lower left panel in \autoref{fig:cfnop} for the face on view at the initial time, t$_0$).  However, the starburst imparts energy and momentum to these clouds, accelerating them radially outwards (upper right panel of \autoref{fig:cfnop}). 

In the simplest scenario, these clouds retain their size as they move outward, and a gap appears between the clouds (see upper right panel in \autoref{fig:cfnop}). The gap allows the background stellar continuum to be transmitted, reducing the covering fraction of the stellar continuum. This is illustrated by the face on view in the lower right panel of \autoref{fig:cfnop}, where background stars become visible in the gaps between the clouds at higher velocities.

This physical picture can be expressed numerically as a ratio of the cloud area to the total surface area at a given radius (r) as \citep{martin09}
\begin{equation}
    \frac{C_f(\mathrm{r})}{C_f(\mathrm{R}_\mathrm{i})} = \frac{\mathrm{A}_\mathrm{c} (\mathrm{r})}{4 \pi \mathrm{r}^2} \frac{4 \pi \mathrm{R}_\mathrm{i}^2}{\mathrm{A}_\mathrm{c}(\mathrm{R}_\mathrm{i})} = \frac{\mathrm{A}_\mathrm{c} (\mathrm{r})}{\mathrm{A}_\mathrm{c} (\mathrm{R}_\mathrm{i})} x^{-2}
    \label{eq:cfchanges}
\end{equation}
Where A$_c$ is the area of the individual cloudlets and $x = \mathrm{r}/\mathrm{R}_\mathrm{i}$. Assuming that the area of the clouds remains constant, the C$_f$ in this scenario scales as x$^{-2}$, significantly steeper than the observed relation of x$^{-0.8}$.

\begin{figure}
\includegraphics[width = 0.5\textwidth]{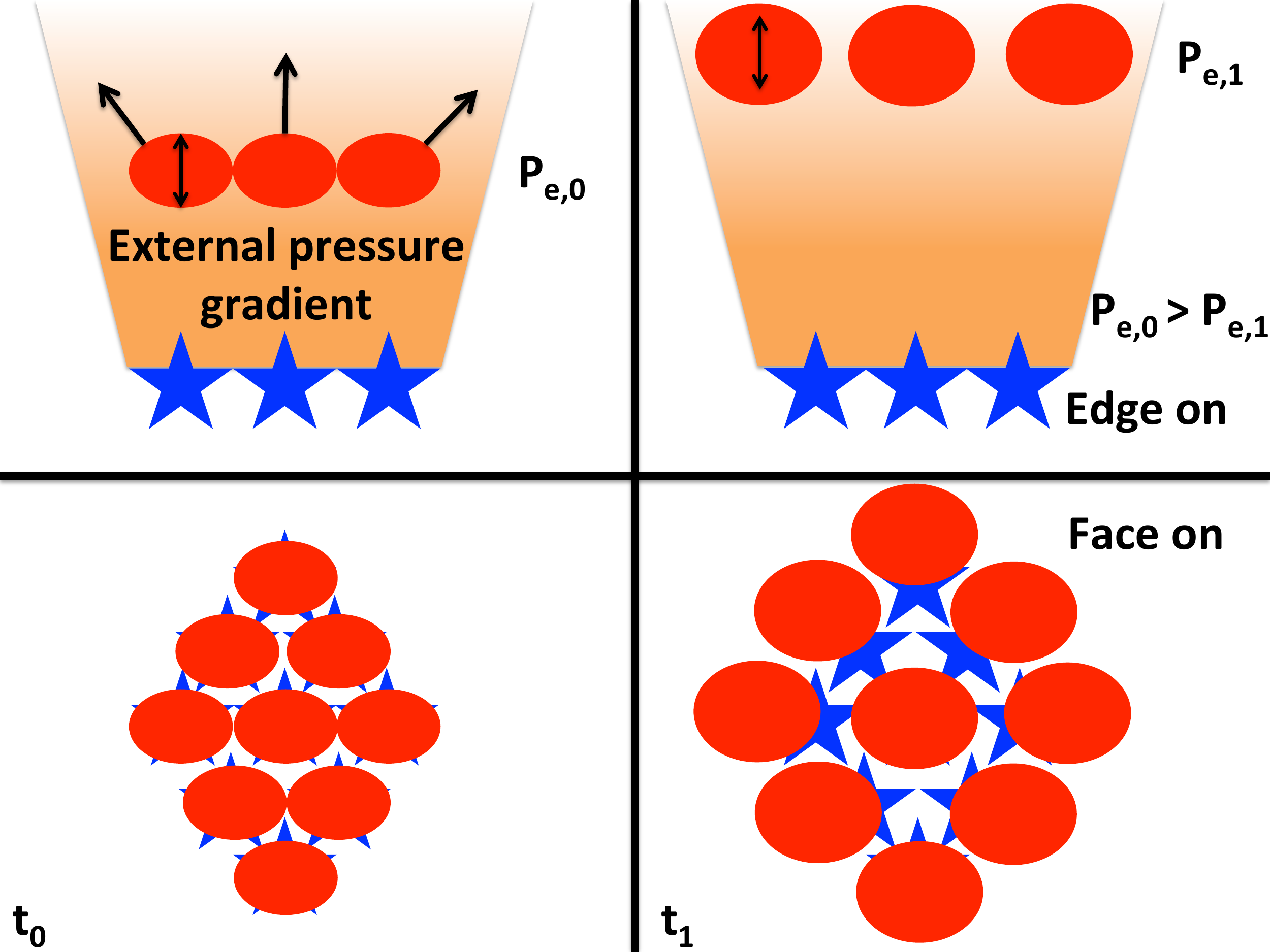}
\caption{Similar to \autoref{fig:cfnop}, but the clouds are in an external pressure gradient (orange shaded region). The initial pressure (P$_\mathrm{e,0}$) is greater than the pressure at t$_1$ (P$_\mathrm{e,1}$). The cloudlets adiabatically expand to remain in pressure equilibrium with the external medium (see the arrow in the upper left cloud demonstrating it's initial diameter). Comparing the bottom right panel with the same panel in \autoref{fig:cfnop}, the expanded clouds now cover more area of the starburst at higher velocities. This slows the drop in C$_f$. The scaling of the C$_f$ with distance depends on how the external medium changes, either isothermally (see \autoref{eq:cftheoryiso}) or adiabatically (see \autoref{eq:cftheoryadia}).}
\label{fig:cfp}
\end{figure}

However, this simple model of static clouds is not the most physical. These cloudlets are likely in pressure equilibrium with an external pressure (P$_\mathrm{e} = \mathrm{P}_\mathrm{c}$, where P$_\mathrm{e}$ is the external pressure and P$_\mathrm{c}$ is the pressure of the cloud), possibly a hot wind (see \autoref{fig:cfp} and the discussion in \autoref{ionstruct}). Changes in P$_\mathrm{e}$ change P$_\mathrm{c}$, and the clouds expand adiabatically to account for these changes as P$_\mathrm{c} \propto V_\mathrm{c}^{-\gamma_\mathrm{c}}$, where V$_\mathrm{c}$ is the volume of the clouds and $\gamma_\mathrm{c}$ is the adiabatic index (5/3 for monatomic ideal gas). The larger cloud volume reduces the gap between the individual cloudlets \citep{martin09}. The lower right panels of \autoref{fig:cfnop} and \autoref{fig:cfp} illustrate that the larger clouds cover more of the background stars than the static clouds do, and C$_f$ falls more slowly with distance. 

Approximating the outflow as spherical clouds provides a relation for A$_\mathrm{c}$ in terms of the volume of the cloudlets as A$_\mathrm{c} \propto \mathrm{V}_\mathrm{c}^{2/3}$. Using this approximation, \autoref{eq:cfchanges}, and the observed C$_f$ (C$_f \propto x^{-0.8}$), the cloud volume changes with distance from the starburst as
\begin{equation}
\frac{\mathrm{V}_\mathrm{c}(\mathrm{r})}{\mathrm{V}_\mathrm{c}(\mathrm{R}_\mathrm{i})} = \left(\frac{\mathrm{A}_\mathrm{c}(\mathrm{r})}{\mathrm{A}_\mathrm{c}(\mathrm{R}_\mathrm{i})}\right)^{3/2} = \left(\frac{C_f(\mathrm{r})}{C_f(\mathrm{R}_\mathrm{i})}x^2\right)^{3/2} =  x^{1.8 \pm 0.4}
\label{eq:volume}
\end{equation}
This demonstrates that the cloud's volume increases with distance from the starburst, but what differential pressure is needed to change the volume as we observe?

Here, we consider two different pressure laws for a mass-conserving external medium: isothermal expansion and adiabatic expansion. If the external pressure changes isothermally with radius then P$_\mathrm{e}(\mathrm{r})/\mathrm{P}_\mathrm{e}(\mathrm{R}_\mathrm{i}) = x^{-2}$. Assuming that the spherical clouds and the isothermal external medium remain in pressure equilibrium (P$_\mathrm{c} = \mathrm{P}_\mathrm{e}$), the pressure of the outflowing clouds changes as
\begin{equation}
\frac{\mathrm{P}_\mathrm{c}(\mathrm{r})}{\mathrm{P}_\mathrm{c}{\mathrm{R}_\mathrm{i}}} = \left(\frac{\mathrm{V}_\mathrm{c}(\mathrm{r})}{\mathrm{V}_\mathrm{c}(\mathrm{R}_\mathrm{i})}\right)^{-\gamma_\mathrm{c}} = \left(\frac{\mathrm{A}_\mathrm{c}(\mathrm{r})}{\mathrm{A}_\mathrm{c}(\mathrm{R}_\mathrm{i})}\right)^{-3\gamma_\mathrm{c}/2} =  x^{-2}
\end{equation}
Solving for $\frac{A_\mathrm{c}(x)}{A_\mathrm{c}(R_i)}$ in terms of x finds that A$_\mathrm{c}$ changes as $x^{4/(3\gamma_\mathrm{c})}$. Placing this into \autoref{eq:cfchanges} gives the C$_f$ scaling of
\begin{equation}
    \frac{C_f(\mathrm{r})}{C_f(\mathrm{R}_\mathrm{i})} = x^{4/(3\gamma_\mathrm{c})} x^{-2} = x^{-1.2}
    \label{eq:cftheoryiso}
\end{equation}
For a $\gamma_\mathrm{c} = 5/3$. If the clouds expand adiabatically while immersed in an isothermal external medium, the covering fraction declines less steeply with height than if the clouds maintain their initial size.

\autoref{eq:cftheoryiso} is still too steep to match the observed scaling. Another possibility is that the external pressure changes adiabatically. In this situation P$_\mathrm{e} \propto x^{-2\gamma_\mathrm{e}}$, where $\gamma_\mathrm{e}$ is the adiabatic index of the external medium. Following a similar process as above, the covering fraction of an adiabatically expanding outflow in pressure equilibrium with an adiabatically expanding external medium is \citep{martin09}
\begin{equation}
        \frac{C_f(\mathrm{r})}{C_f(\mathrm{R}_\mathrm{i})} = x^{4\gamma_\mathrm{e}/(3\gamma_\mathrm{c}) -2} = x^{-2/3}
        \label{eq:cftheoryadia}
\end{equation}
where we have assumed that $\gamma_\mathrm{e}$ and $\gamma_\mathrm{c}$ are equal to each other (i.e. both the clouds and external medium are monatomic ideal gases). Interestingly, since diatomic gas has more degrees of freedom than monatomic gas, diatomic gas requires a larger energy change to remain in pressure equilibrium, as seen by the smaller $\gamma_\mathrm{c}$ of diatomic gas (7/5). Using \autoref{eq:cftheoryadia}, we find that molecular clouds expand more rapidly, and consequently have a more gradually declining C$_f$ with distance, as x$^{-0.4}$. This has an important consequence for the survival of molecular clouds in a hot medium. By increasing the radius of the cloud, it takes longer for a shock wave to propagate across the cloud and dissociate the molecular gas  \citep{klein, scannapieco15}. This may extend the lifetime of molecular gas entrained in galactic outflows, while the increased  C$_f$ may explain the presence of molecular gas at high velocities \citep{matsushita2000, sakamoto, leroy15}.

In \autoref{beta}, we find that the covering fraction scales as $C_f~=~1.0~x^{-0.8 \pm 0.2}$, in agreement with the adiabatic expansion of cloudlets in an adiabatically expanding external medium (\autoref{eq:cftheoryadia}). However, it is an interesting theoretical question whether clouds can remain in pressure equilibrium with an external medium and still be accelerated. Ram pressure introduces a secondary pressure term, but if the clouds are shielded from direct interaction with the hot medium then the pressure of the cloud is set by the external thermal pressure \citep{chevalier}. Further, ablation and ram pressure stripping create a fractal distribution into smaller cloudlets \citep{klein, scannapieco15}, which may produce the observed change in covering fraction. The velocity scaling of the covering fraction provides further constraints for simulations of the acceleration of clouds. The acceleration of the cloudlets out of the star forming region decreases C$_f$, but how are the clouds accelerated?

\subsection{Velocity Law}
\label{velocity}

\begin{figure}
\includegraphics[width = 0.5\textwidth]{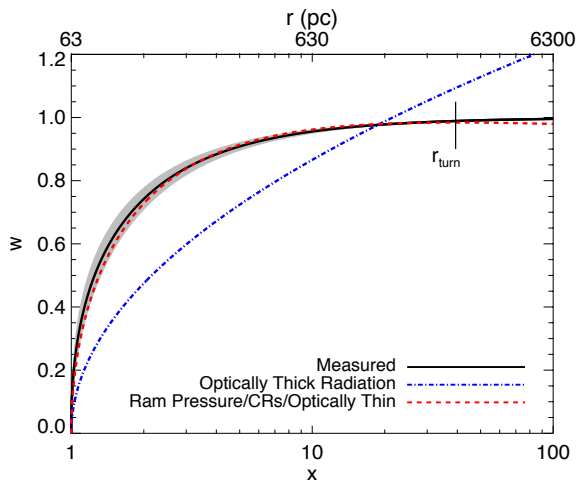}
\caption{Normalized velocity ($w = \mathrm{v}/\mathrm{v}_\infty$; $\mathrm{v}_\infty = -399$~\kmsp) profile with normalized radius (x = r/R$_\mathrm{i}$; R$_\mathrm{i}$ = 63.4~pc, see \autoref{eq:r}). The solid black curve represents the measured $\beta$ velocity law from \autoref{beta}, while the grey shaded region corresponds to the fitted error on $\beta$. The blue dot-dashed and red dashed lines represent theoretical velocity profiles of optically thick radiation pressure and ram pressure/cosmic ray/optically thin radiation pressure (any r$^{-2}$ force in opposition with gravity) driven outflows, respectively. The upper x-axis is given in terms of pc, as measured in \autoref{ionstruct}. At 2.5~kpc the r$^{-2}$ profile begins to decelerate, and this radius (r$_\text{turn}$) is marked by a vertical line.}
\label{fig:vellaw}
\end{figure}

The velocity law describes the acceleration of the outflowing clouds with distance. In \autoref{beta} we find a $\beta$ velocity profile of $\mathrm{v(r)}~=~\mathrm{v}_\infty~(1~-~\frac{\mathrm{R}_\mathrm{i}}{\mathrm{r}})^{0.43 \pm 0.07}$. The outflow initially accelerates rapidly, but the acceleration moderates at large radii (see the black line in \autoref{fig:vellaw}). Using the initial radius calculated in \autoref{ionstruct}, the outflow reaches 50\% (90\%) of the maximum velocity in 79~pc (291~pc). Acceleration much beyond this is not well constrained by the data. 

Outflows typically have a "saw-tooth" line profile \citep{weiner}, where the red side of the profile sharply declines and the blue side rises gradually. The velocity and C$_f$ laws produce these profiles. Initially, the velocity sharply increases over a short distance, keeping the C$_f$ near unity because C$_f$ scales as x$^{-0.82}$. Meanwhile, at higher velocities the clouds travel larger distances per velocity interval, forcing the high-velocity clouds to expand to remain in pressure equilibrium with the adiabatically expanding external medium. This moderates the decline in C$_f$, and produces the gentler rising blue portion of the profile. 

The observed \siivp~1402\AA\ outflows do not escape the galactic potential. Following \citet{heckman2000}, the escape velocity is no more than three times the circular velocity \citep[136~\kmsp;][]{chisholm15}, therefore the outflow velocity needs to exceed 651~\kms for the clouds to escape the potential. There are two possible explanations for why we do not observe gas escaping the potential: (1) the outflow does eventually exceed the escape velocity but the \siiv density drops below the detection limits at high velocities and we do not observe the escaping gas  (see \autoref{density}) or (2) the outflow does not actually escape the galaxy, but rather recycles back into the disc as a galactic fountain \citep{shapiro}. The latter is consistent with \citet{chisholm15}, which finds most galaxies with log(\mstarp/M$_\odot$) greater than 10.5 cannot drive outflows faster than their escape velocity, unless they are undergoing a merger. Observations probing lower density gas, such as Lyman-$\alpha$, may constrain whether these outflows are capable of escaping the gravitational potential. 

Now we study how these outflows are accelerated. There are numerous theoretical ways to accelerate galactic outflows: radiation pressure on dust grains \citep{thompson05, murray05}, cosmic rays \citep{everett, socrates}, and ram pressure of a hot wind on the cloudlets \citep{murray05, cooper08, fujita09, martin09}. Below, we use analytical expressions for how the velocity profile evolves radially to explore which mechanisms could accelerate these outflows. In each case, we give the scaling of the normalized velocity (w = v/v$_\infty$; where v$_\infty$ is the maximum velocity) with the normalized radius (x = r/R$_\mathrm{i}$; where R$_\mathrm{i}$ is the initial radius). We then scale the analytical expressions to the observed relations to determine the plausibility of each mechanism. In \autoref{velsum} we summarize the implications for the theoretical profiles.

\subsubsection{Optically Thick Radiation Pressure}
\label{radiation}

Radiation pressure is an attractive way to drive outflows in dusty, vigorously star-forming galaxies: the high luminosity provides a large momentum source \citep{murray05}, while the large dust optical depth scatters photons multiple times \citep{thompson05}. \citet{murray05} give the radial scaling of the velocity for optically thick radiation pressure as:
\begin{equation}
w = \sqrt{\frac{4\sigma^2}{\mathrm{v}_\infty^2} (\frac{L}{L_\mathrm{E}}-1)~\text{ln}(x)}
\label{eq:radvel}
\end{equation}
where $\sigma$ is the velocity dispersion of the galaxy, L is the luminosity of the galaxy, and L$_\mathrm{E}$ is the Eddington luminosity. We fit for the constant value that best matches the observed $\beta$-law using MPFIT \citep{mpfit}, while excluding velocities less than 0.2. The fit is shown by the blue dot-dashed line in \autoref{fig:vellaw}. The optically thick radiation model poorly matches the observations.

\subsubsection{An r$^{-2}$ Force}
\label{ram}

A second appealing driving mechanism is ram pressure. Supernovae thermalize ambient gas into a hot wind, which expands adiabatically out of the star forming region. The hot wind imparts a ram pressure force on the clouds, which depends on the speed and density of the hot wind and the area of the cloud as F$_\mathrm{ram} = \rho_\mathrm{h} \mathrm{v}_\mathrm{h}^2 \mathrm{A}_\mathrm{c}$ \citep{klein, scannapieco15}. F$_\mathrm{ram} \propto x^{-2}$ in a simplified case of a mass conserving, adiabatically expanding hot wind.  Ram pressure driving is appealing because the C$_f$ scaling is consistent with clouds being in pressure equilibrium with a adiabatically expanding external medium (\autoref{cf}), a situation that could lead to ram pressure driving. Including the effects of gravity, \citet{murray05} calculate the velocity profile of a ram pressure driven outflow as
\begin{equation}
w(x) = \sqrt{\frac{v_\mathrm{c}^2}{\mathrm{v}_\infty^2} (1-\frac{1}{x}) -\frac{4\sigma^2}{\mathrm{v}_\infty^2} \mathrm{ln}(x)}
\label{eq:ramvel}
\end{equation}
Where v$_c$, the characteristic velocity clouds reach before gravity dominates, is defined as v$_\mathrm{c} = \frac{3\dot{M}_\mathrm{h} \mathrm{V}_\mathrm{h}}{8\pi \rho_\mathrm{c} \mathrm{R}_\mathrm{c} \mathrm{R}_\mathrm{i}}$. Defining A as v$_\mathrm{c}^2/\mathrm{v}_\infty^2$ and B as $4\sigma^2/\mathrm{v}_\infty^2$, we fit for the values of A (fitted value of 1.10) and B (fitted value of 0.03) that match the observed $\beta$ velocity law, as shown by the red line in \autoref{fig:vellaw}. 

Similarly, shockwaves from supernovae accelerate cosmic rays (CR). These relativistic particles stream out of the star forming regions along the magnetic fields and exchange momentum with magnetized plasma. Cosmic rays have various advantages over radiation pressure and ram pressure, including: CRs interact multiple times with the gas, magnetic fields confine CRs to the galaxy making it difficult for CRs to escape without imparting momentum, and CR feedback is independent of the distribution of ISM gas \citep{everett, socrates}. \citet{socrates} derive the force imparted by cosmic rays as
\begin{equation}
    F_\mathrm{CR} = \frac{\kappa_\mathrm{CR}}{c} \frac{L_\mathrm{CR}}{4 \pi r^2} 
\end{equation}
where $\kappa_\mathrm{CR}$ is the cosmic ray opacity, and L$_\mathrm{CR}$ is the cosmic ray luminosity. This force produces a similar velocity scaling as the ram pressure scaling in \autoref{eq:ramvel}. 

Additionally, the scaling of the ram pressure and CR velocity profiles are similar to an optically thin radiation profile \citep{murray05}. In fact, \autoref{eq:ramvel} is a general form for any r$^{-2}$ force that opposes gravity. Therefore, the red dashed line in \autoref{fig:vellaw} corresponds to outflows driven by any r$^{-2}$ force (ram pressure, cosmic rays, or optically thin radiation pressure), and we cannot distinguish these mechanisms from the measured velocity profile.    

\subsubsection{Velocity Summary}
\label{velsum}
The ram pressure, cosmic rays, and optically thin radiation pressure velocity profile (any r$^{-2}$ force) matches the observed velocity relation for all observed velocities, while optically thick radiation pressure poorly matches the observed velocity profile (\autoref{fig:vellaw}). In \autoref{fig:vellaw}, the r$^{-2}$ profile begins to decelerate at a radius of 2.5~kpc (r$_\text{turn}$; marked by the vertical line in \autoref{fig:vellaw}). However, at these large radii the outflow is now diffuse, and the deceleration of the outflow is challenging to study with the \siiv line profile.

\subsection{Density Law}
\label{density}
\begin{figure}
\includegraphics[width = 0.5\textwidth]{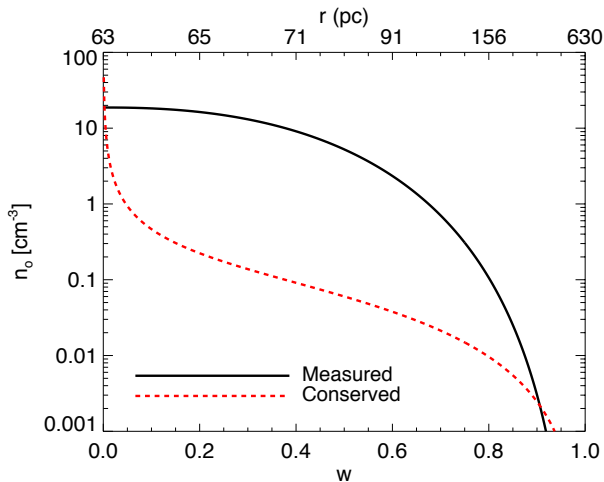}
\caption{The measured outflow Hydrogen density scaling with normalized velocity ($w=v/v_\infty$) on the lower x-axis and radius on the upper x-axis. The density does not change appreciably at low velocities because the velocity rapidly increases over a short distance from the starburst (see \autoref{fig:vellaw}). At higher velocities the $\beta$-law flattens, and increases in velocity are accompanied by larger increases in radius. The red dashed line shows the density approximation of a mass-conserving flow with the observed velocity law (\autoref{eq:masscon}). In a mass conserving flow, the density rapidly decreases at low velocity as the outflow is accelerated. Note that the density profile is only fit to the observations for $w > 0.2$. The radius formally extends to infinity, but the last tick corresponds to the radius of r/R$_\mathrm{i}$ = 10.}
\label{fig:denlaw}
\end{figure}

In \autoref{beta} we find the density to scale with the normalized radius as $\mathrm{n}(x) \propto x^{-5.72}$. In \autoref{fig:denlaw} we show the measured relation for the outflow density with velocity. At low velocities, the density remains nearly constant because the outflow accelerates over a short distance \citep{steidel10}, but at higher velocities the density precipitously drops. Stronger transitions probe lower densities than weaker transitions, causing stronger transitions to probe higher velocities than the weaker transitions. Therefore, stronger transitions have larger measured terminal velocities and larger line widths (because they probe a wider velocity distribution) than the weaker transitions \citep[see \autoref{tab:obs};][]{grimes09, chisholm16}. 

This density-velocity scaling implies that the outflow traced by \siiv does not conserve mass. Fitting the $\tau$ distribution with a mass-conserving flow produces a poor fit because the $\tau$ distribution declines sharply at high velocity. In the continuity equation, the mass flux is conserved, such that the density scales as
\begin{equation}
    \mathrm{n}(w) \propto \frac{n_\mathrm{0}}{x^2v} \propto \frac{n_\mathrm{0}}{v_\infty}\frac{(1-w^{1/\beta})^2}{w}
    \label{eq:masscon}
\end{equation}
In \autoref{fig:denlaw} we compare the mass-conserving density profile (red dashed line) with the observed profile. Compared to the observed density profile, the density of a mass-conserving flow rapidly decreases at low velocities as the outflow accelerates over small distances, while at higher velocities the conserved density decreases more gradually because the velocity gradient is flatter (see \autoref{fig:vellaw}). 

If the \siiv flow is not mass-conserving, what happens to the \siiv gas? The density could decrease by: (1) increasing the volume of the outflowing clouds while keeping the number of \siiv ions constant or (2)  decreasing the number of \siiv ions in the clouds. \autoref{eq:volume} shows that the volume of the outflowing clouds scales as x$^{1.77}$, implying that the number of \siiv ions in the outflow must decrease as x$^{-3.95}$ to satisfy the observed density relation. One possibility is that the outflowing clouds lose mass through ablation, ram pressure stripping, or conduction from the hot wind \citep{klein, scannapieco15, zhang, bruggen}. In this scenario, the hot wind rapidly destroys the clouds and incorporates them into the hot wind, making the mass undetectable through \siiv absorption at higher velocities.

A steep density scaling relation is also seen in the nearby starburst M~82. \citet{leroy15} probe the surface density profiles of the molecular, neutral and ionized phases of the outflow from M~82. They find that only the H~{\sc I} follows an n$ \propto \mathrm{r}^{-2}$ density law largely because of the large amount of tidal material present at large radii.  However, other surface density profiles along the minor axis decrease more rapidly. The 70~$\mu$m emission, a tracer of warm dust, and CO emission, a tracer of diffuse molecular gas, scale with the distance from the starburst as n$ \propto \mathrm{r}^{-4}$ \citep{leroy15}, roughly consistent, with the density scaling found here. The authors use the divergence of the density profile to argue that the outflow of M82 is a galactic fountain, with gas leaving the outflow and recycling back into the disk.

\subsection{Mass Outflow Rate}
\label{mout}
\begin{figure}
\includegraphics[width = 0.5\textwidth]{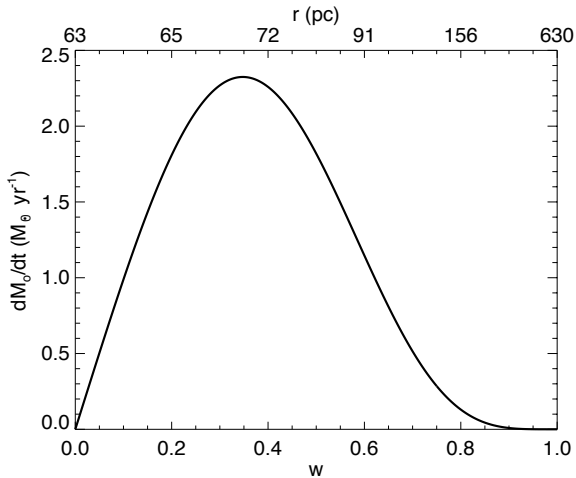}
\caption{The mass outflow rate (\mout = dM$_\mathrm{o}$/dt) of NGC~6090 with normalized velocity ($w = v/v_\infty$) on the lower axis, and radius, in pc, on the upper axis. The \mout is calculated using \autoref{eq:mout}, the fitted velocity profile, density profile, covering fraction profile, metallicity, and ionization structure of the outflow. The only assumed parameter is that the outflow has a full opening angle of 140$^{\circ}$ ($\Omega = 3.11\pi$~steradians). The peak \mout is 2.32~M$_\odot$~yr$^{-1}$. Note that the radius formally extends to infinity, and the last tick corresponds to the radius of x = 10.}
\label{fig:mout}
\end{figure}

Finally, we combine all of the derived relations to measure the mass outflow rate ($\dot{M}_o$ = dM$_o$/dt) of the photoionized galacitc outflow. The photoionized outflow is a single component of the outflow, and other, unexplored, phases likely contribute to the total mass outflow rate. Since the hot phase is such low density, this photoionized phases likely dominates the mass budget of the outflow.

Typically the mass outflow rate is calculated assuming the outflow is in a thin spherical shell as
\begin{equation}
    \dot{M}_\mathrm{o} = \Omega {C}_{f} \mu \mathrm{m}_\mathrm{p} \mathrm{N}_\mathrm{H} \mathrm{r} \mathrm{v}_\mathrm{cen}
\end{equation}
where $\mu m_\mathrm{p}$ is 1.4 times the mass of the proton, $\Omega$ is the angular covering fraction of the wind and v$_\mathrm{cen}$ is the centroid velocity of the outflow (132~\kmsp). Previous studies assume a radius of 5~kpc, a full opening angle of 140$^\circ$ (and $\Omega = 3.11\pi$~steradians), solar metallicity, and no ionization correction \citep{rupke2005b, martin2005, weiner, rubin13}. Using the values from the \siiv line, and these assumptions, we derive a mass outflow rate for NGC~6090 of 7.32~M$_\odot$~yr$^{-1}$. However, previous studies typically use absorption lines of cooler gas like Na~{\sc I}, Mg~{\sc II} and Fe~{\sc II} \citep{rupke2005b, martin2005, weiner, martin09, rubin13}. If we use the \siii 1304 column density, a similar ionization potential to the previous outflow tracers, \mout rises to 22.6~M$_\odot$~yr$^{-1}$.

However, with our tightly constrained physical model of the outflow, now we only have to assume an angular covering fraction to calculate the mass outflow rate. Above, we derive a radius, a metallicity, and an ionization fraction for the outflows (see \autoref{tab:ion} for values). If we use the radius of peak optical depth (R$_\mathrm{p}$ = 72.2~pc, at a w of 0.41), we calculate a total mass outflow rate of 0.81~M$_\odot$~yr$^{-1}$, 28 times lower than the \mout calculated using \siiip.

While this calculation uses many of our derived values, it ignores the evolution of these quantities with velocity. In \autoref{velocity} we find that the outflow is accelerated over short distances, in \autoref{cf} we find that the covering fraction decreases at large velocities, and in \autoref{density} we find that the outflow is not a mass conserving flow, rather it evolves as x$^{-5.7}$. All of these impact how mass is distributed in velocity space. Using the scaling of the covering fraction, radius, and density with velocity, we define the mass outflow rate per velocity as
\begin{equation}
\begin{aligned}
    \dot{M}_\mathrm{o}(\mathrm{r}) &= \Omega C_{f}(\mathrm{r}) \mathrm{v}(\mathrm{r}) \rho(\mathrm{r}) \mathrm{r}^2 \\
    \dot{M}_\mathrm{o}(w)&= \Omega C_f(\mathrm{R}_\mathrm{i}) \mathrm{v}_\infty \mu \mathrm{m}_\mathrm{p}  \mathrm{n}_\mathrm{0} \mathrm{R}_\mathrm{i}^2 \frac{w}{(1-w^{1/\beta})^{2+\gamma+\alpha}} \\
    &= 10.1~\text{M}_\odot~\text{yr}^{-1} \frac{w}{(1-w^{1/\beta})^{2+\gamma+\alpha}}
    \label{eq:mout}
\end{aligned}
\end{equation}
where we use the values from \autoref{tab:beta} and \autoref{tab:ion} for the constants and the exponents, and the only assumed parameter is that $\Omega$ is 3.11$\pi$~steradians. The \mout relation is not constant with velocity, as typically assumed (\autoref{fig:mout}): \mout increases rapidly at low velocities as the outflow accelerates, and declines at high velocities as the density and covering fraction of the outflow decline (\autoref{fig:vellaw} and \autoref{fig:denlaw}). 

At w of 0.35, the \mout relation peaks at a value of 2.3~M$_\odot$~yr$^{-1}$. This is a factor of ten times smaller than the \siii value calculated with previous assumptions for geometry, ionization fractions, and metallicities. Combining \mout with the derived outflow metallicity, we find that the maximum metal outflow rate is 0.07~M$_\odot$~yr$^{-1}$. This may constrain the impact of outflows on the mass-metallicity relationship \citep[see above;][]{tremonti04, finlator08, peeples11,andrews, zahid, creasey2015, christensen}.

It is typical to normalize the mass outflow rate by the star formation rate to produce the "mass-loading" factor \citep[$\eta=\dot{M}_o/SFR$;][]{rupke2005b}. The global $\eta$ is 0.09. This $\eta$ is smaller than recent measurements from \citet{heckman2015}, who find an $\eta$ near 0.3 for galaxies with SFRs near 30~M$_\odot$~yr$^{-1}$. However, \citet{heckman2015} use the mean ISM metallicity of the sample (0.5~Z$_\odot$), do not calculate ionization fractions, and use a multiple of the half-light radius (typically 0.5-1~kpc) as the radius of the outflow. Other studies make similar assumptions about the metallicity and the ionization state, but assume constant outflow radii between 1-5~kpc \citep{rupkee2005, martin12, rubin13}. These studies typically find $\eta$ between 0.02 and 1 for SFRs and \mstar similar to NGC~6090. The method outlined here does not rely on these uncertain assumptions and allows for these quantities to vary from galaxy to galaxy, reducing the scatter in the measurements.

The relatively small inner outflow radius implies that the local SFR may be more important than the global SFR. In \citet{chisholm15} we calculate the fraction of the total {\it GALEX} UV flux within the COS aperture to be 22\%, leading to a SFR within the COS aperture of 5.55~M$_\odot$~yr$^{-1}$, however this does not account for the significant spatial variation in the IR flux and is only a crude approximation for the local SFR. Using this local SFR we calculate a local $\eta$ within the COS aperture of 0.42.

The local and global $\eta$ values are a factor of 5 and 22 times lower than the redshift independent $\eta$ predicted by the FIRE simulation \citep{hopkins12b,fire,muratov}, although the high-mass FIRE galaxies at redshift 0 only have upper limits for $\eta$ \citep{muratov}. \citet{hayward} develop an analytical model that uses the momentum from stellar feedback to regulate star formation and drive outflows. A requirement of this model is that $\eta$ depends on the gas fraction and the stellar mass of the host galaxy. Making approximations for these scalings with redshift, the analytical model predicts the mass-loading for a log(\mstarp/M$_\odot$) galaxy of 10.7 at $z\approx0.03$ to be 0.05, in rough agreement with the observed global $\eta$. While the outflow from NGC~6090 is weak compared to the SFR, \citet{hayward} predict that 10$^9$~M$_\odot$ galaxies drive outflows with an $\eta$ of 10 at $z\sim0$. In a future paper we will explore the scaling of \mout and $\eta$ with host galaxy properties to better constrain these types of studies.

\begin{figure}
\includegraphics[width = 0.5\textwidth]{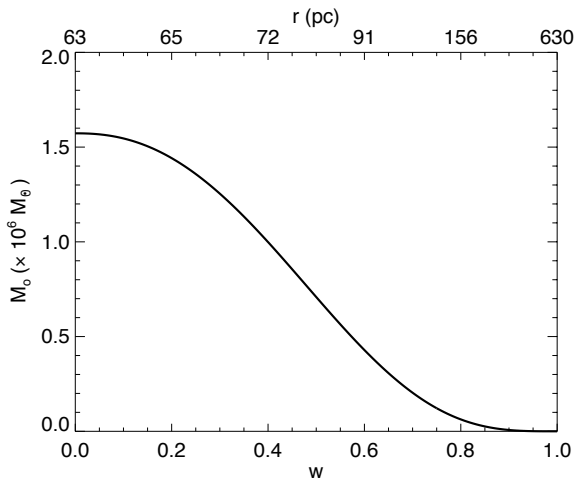}
\caption{The total mass within the outflow at each velocity, as given by \autoref{eq:m}. The total integrated outflow mass is $7.5 \times 10^5$~M$_\odot$.}
\label{fig:m}
\end{figure}

The total mass in the outflow (M$_o$) at a particular velocity is given by 
\begin{equation}
\begin{aligned}
    M_\mathrm{o} (w) &= \Omega C_f(\mathrm{r}) \mu \mathrm{m}_\mathrm{p} \mathrm{n}_\mathrm{0} \mathrm{R}_\mathrm{i}^3 \left(\frac{1}{1-w^{1/\beta}}\right)^{3+\alpha+\gamma} \\
    & = 1.5 \times 10^{6} M_\odot  \left(\frac{1}{1-w^{1/\beta}}\right)^{3+\alpha+\gamma}
    \label{eq:m}
\end{aligned}
\end{equation}
This relation is shown in \autoref{fig:m}. Like the density, the total mass in the \siiv outflow declines with velocity. The integrated mass between w of 0 and 1 is $7.5 \times 10^{5}~M_\odot$. While the current mass of outflowing gas is significantly lower than the measured H~{\sc I} mass of 10$^{10.2}$~M$_\odot$ \citep{vandriel}, if \mout remains constant over the $\sim1$~Gyr time scale of the merger, than the outflow will process (and enrich) 15\% of the observed H~{\sc I} in NGC~6090, possibly leading to the metal enriched gas seen in the halos of galaxies \citep{tumlinson, werk, peeples14, wakker2015}. 

\section{CONCLUSION}
Here we measure a physically motivated mass outflow rate (\moutp) of the nearby starburst NGC~6090. To calculate \moutp, we first fit the optical depth with a Sobolev optical depth, and the covering fraction with a radial power-law  (see \autoref{eq:cfbeta} and \autoref{fig:beta}). We then calculate the ionization corrections and the metallicity of the outflow using a Bayesian analysis, {\sc CLOUDY} models, and the measured column densities (see \autoref{tab:ion} and \autoref{fig:ionpdfs}). The main results of this study are:
\begin{enumerate}
    \item The ionization model estimates the metallicity (1.61~Z$_\odot$), density (18.73~cm$^{-3}$) and the ionization parameter (log(U) = -1.85) of the outflow. The estimated H column density is consistent with that derived from the UV continuum extinction. The outflow is at least as metal enriched as the ISM of the host galaxy (see \autoref{ionstruct}).
    \item Using the absorption line profile and ionization models, we determine that the inner edge (R$_\mathrm{i}$) of the outflow is 63~pc from the starburst (see \autoref{eq:r}), consistent with the absorption arising from the shredded blastwaves of supernovae remnants. Most of the outflow is constrained within 300~pc of the starburst (\autoref{ionstruct}). 
    \item The covering fraction scales as C$_f = 1.0 (\mathrm{r}/\mathrm{R}_\mathrm{i})^{-0.8\pm0.2}$ (\autoref{tab:beta}). This scaling relation is consistent with predictions of clouds in pressure equilibrium with an adiabatically expanding external medium (\autoref{cf}). This warrents further study of whether outflowing clouds remain in pressure equilibrium with an external medium, and if a fractal cloud distribution could produce a similar scaling relation.
    \item The \siiv outflow velocity scales with radius as v$~=~\mathrm{v}_\infty (1~-~\mathrm{R}_\mathrm{i}/\mathrm{r})^{0.43 \pm 0.07}$ (\autoref{fig:vellaw} and \autoref{tab:beta}), where v$_\infty$ is the maximum velocity. We compare this velocity profile to models of different driving mechanisms, and find that an r$^{-2}$ force law matches the observed profile (see \autoref{fig:vellaw} and \autoref{velocity}).
    \item The outflow density decreases with radius as n$ \propto \mathrm{r}^{-5.7 \pm 1.5}$ (\autoref{fig:denlaw}), which declines more rapidly than a mass conserving flow (\autoref{density}). This rapid density reduction could be due to interactions between the outflowing clouds and a hotter wind. 
    \item Combining all of our measurements, we derive a maximum  mass outflow rate (\moutp) of 2.3~M$_\odot$~yr$^{-1}$ (see \autoref{fig:mout}). The mass-loading factor (mass outflow rate divided by star formation rate) is 0.09. The \mout is a factor of 10 lower than the \mout calculated using common assumptions for ionization state, metallicity, and geometry (see \autoref{mout}). 
\end{enumerate}
In future work, we will continue this analysis for a larger sample of star forming galaxies, studying how the outflow properties (metallicity, initial radius, mass outflow rate) scale with host galaxy properties. These scaling relations will constrain future models of galaxy evolution.

\section*{Acknowledgments}

We thank the anonymous referee for constructive comments that strengthened the paper. 
Joseph Cassinelli inspired this work with helpful conversations and notes.
We thank Bart Wakker for help with the data reduction and discussions on the analysis.

Support for program 13239 was provided by NASA through a grant from the Space Telescope Science Institute, which is operated by the Association of Universities for Research in Astronomy, Inc., under NASA contract NAS 5-26555. Some of the data presented in this paper were obtained from the Mikulski Archive for Space Telescopes (MAST). STScI is operated by the Association of Universities for Research in Astronomy, Inc., under NASA contract NAS 5-26555.  

All of the HST data presented in this paper were obtained from the Mikulski Archive for Space Telescopes (MAST). STScI is operated by the Association of Universities for Research in Astronomy, Inc., under NASA contract NAS5-26555. Support for MAST for non-HST data is provided by the NASA Office of Space Science via grant NNX09AF08G and by other grants and contracts. 

Funding for the Sloan Digital Sky Survey (SDSS) has been provided by the Alfred P. Sloan Foundation, the Participating Institutions, the National Aeronautics and Space Administration, the National Science Foundation, the U.S. Department of Energy, the Japanese Monbukagakusho, and the Max Planck Society. The SDSS Web site is http:\\www.sdss.org/. The SDSS is managed by the Astrophysical Research Consortium (ARC) for the Participating Institutions. The Participating Institutions are The University of Chicago, Fermilab, the Institute for Advanced Study, the Japan Participation Group, The Johns Hopkins University, Los Alamos National Laboratory, the Max-Planck-Institute for Astronomy (MPIA), the Max-Planck-Institute for Astrophysics (MPA), New Mexico State University, University of Pittsburgh, Princeton University, the United States Naval Observatory, and the University of Washington.

\bibliographystyle{mnras}
\bibliography{ngc6090}

\bsp	% typesetting comment
\label{lastpage}

\end{document}